\documentclass[manuscript,screen,review=false]{acmart}

\AtBeginDocument{%
  }

\usepackage{ulem}
\usepackage{latexsym}
\usepackage{amsmath}
\usepackage{amsthm}
\usepackage{graphicx}
\usepackage{enumerate, color}
\usepackage{multirow,bigdelim}
\usepackage{arydshln}
\usepackage{mathtools}
\usepackage{hyperref}
\usepackage{algorithm}
\usepackage{algorithmic}
\usepackage{rotating}
\usepackage{subfigure}
\usepackage{float}
\theoremstyle{definition}
\newtheorem{theorem}{Theorem}

\newtheorem{definition}[theorem]{Definition}

\newcommand{\tabincell}[2]{\begin{tabular}{@{}#1@{}}#2\end{tabular}}  
\usepackage{threeparttable}

\usepackage{adjustbox}
\usepackage{lipsum} 
\usepackage{pdflscape}
\usepackage{enumitem}
\setcopyright{acmlicensed}
\copyrightyear{2018}
\acmYear{2018}
\acmDOI{XXXXXXX.XXXXXXX}

\acmConference[Conference acronym 'XX]{Make sure to enter the correct
  conference title from your rights confirmation emai}{June 03--05,
  2018}{Woodstock, NY}
\acmISBN{978-1-4503-XXXX-X/18/06}

\begin{document}

\title{Tabular Data Synthesis with Differential Privacy: A Survey}

\author{Mengmeng Yang}
\affiliation{%
  \institution{CSIRO, Data61}
  \city{Melbourne}
  \country{Australia}}
\email{mengmeng.yang@data61.csiro.au}

\author{Chi-Hung Chi, Kwok-Yan Lam}
\affiliation{%
  \institution{Nanyang Technological University}
  \country{Singapore}}
\email{chihung.chi@ntu.edu.sg}

\author{Jie Feng}
\affiliation{%
  \institution{Xidian University}
  \city{Xi'an}
  \country{China}}
\email{fengjie@xidian.edu.cn}
\authornote{Participated in the work during the visit to NTU.}

\author{Taolin Guo}
\affiliation{%
  \institution{Chongqing Normal University}
  \city{Chongqing}
  \country{China}}
\email{tguo@cqnu.edu.cn}

\author{Wei Ni}
\affiliation{%
  \institution{CSIRO, Data61}
  \city{Sydney}
  \country{Australia}}
\email{wei.ni@data61.csiro.au}

\renewcommand{\shortauthors}{Yang et al.}

\begin{abstract}

Data sharing is a prerequisite for collaborative innovation, enabling organizations to leverage diverse datasets for deeper insights. In real-world applications like FinTech and Smart Manufacturing, transactional data, often in tabular form, are generated and analyzed for insight generation. However, such datasets typically contain sensitive personal/business information, raising privacy concerns and regulatory risks. Data synthesis tackles this by generating artificial datasets that preserve the statistical characteristics of real data, removing direct links to individuals. However, attackers can still infer sensitive information using background knowledge. Differential privacy offers a solution by providing provable and quantifiable privacy protection. Consequently, differentially private data synthesis has emerged as a promising approach to privacy-aware data sharing. This paper provides a comprehensive overview of existing differentially private tabular data synthesis methods, highlighting the unique challenges of each generation model for generating tabular data under differential privacy constraints. We classify the methods into statistical and deep learning-based approaches based on their generation models, discussing them in both centralized and distributed environments. We evaluate and compare those methods within each category, highlighting their strengths and weaknesses in terms of utility, privacy, and computational complexity. %Additionally, we present evaluation methods and highlight research gaps, proposing future research directions. 
Additionally, we present and discuss various evaluation methods for assessing the quality of the synthesized data, identify research gaps in the field and directions for future research. 
\end{abstract}



\received{20 February 2007}
\received[revised]{12 March 2009}
\received[accepted]{5 June 2009}

\maketitle

\section{Introduction}

Data sharing is essential as it drives innovative collaboration and enables informed decision-making across various domains. 
Numerous public data-sharing platforms, including Kaggle \cite{kaggle}, Data.gov \cite{datagov}, and the UCI repository \cite{UCI}, offer access to extensive datasets, with the primary goal of facilitating knowledge discovery and advancement. In most applications, such as FinTech and Smart Manufacturing, these datasets are represented in tabular form, given their structured nature and widespread applicability across different fields. However, it is important to note that these datasets often contain sensitive personal/business data, which can raise significant privacy concerns. In addition, due to evolving privacy regulations, exemplified by recent legislation like the AI Act \cite{AIact}, there is a heightened need for innovative methods for data sharing that protect individual privacy while enabling meaningful data analysis. 

Data synthesis has been attracting growing attention due to its unique ability to generate synthetic data based on statistical information without being linked to specific individuals or identities. However, it is important to note that while synthetic data offers privacy protection, several studies \cite{dinur2003revealing,garfinkel2019understanding} have shown that the attacker can still potentially infer sensitive information about users. 
For example, Jordon et al. \cite{jordon2022synthetic} show that \textit{``Synthetic data can leak information about the data it was derived from and is vulnerable to privacy attacks."} Moreover, Stadler et al. \cite{stadler2020synthetic} have demonstrated that generative models trained without privacy safeguards offer limited defence against inference attacks when compared to the alternative of directly sharing the original data. 

A cutting-edge solution involves integrating provable privacy measures, such as differential privacy (DP), into the synthetic data generation process. Differential privacy aims to ensure that the information derived from the released synthetic data remains nearly identical, whether or not specific individuals were part of the original datasets, thus effectively preventing the inference of personal information. Importantly, it does not rely on assumptions about the capabilities of potential attackers, providing robust privacy protection even in the presence of adversaries with significant background knowledge and resources \cite{yang2022differentially}. The United States National Institute of Standards and Technology (NIST) has been instrumental in championing data sharing and privacy protection. In 2018, NIST organized the ``Differential Privacy Synthetic Data Challenge" \cite{NISTchallenge}, a competition dedicated to advancing the field of differential privacy for generating synthetic data that retains the statistical characteristics of actual data while safeguarding individual privacy. The challenge highlighted the growing importance of balancing the need to share valuable insights with the necessity of protecting personal information, making differentially private data synthesis a promising research focus. 
%Striking this balance between sharing valuable insights and safeguarding personal information is crucial in harnessing the full potential of data for the benefit of society and governance. 

% In this paper, we conduct a comprehensive review of existing differential privacy methods for synthesizing tabular data. %Tabular data, renowned for its structured layout of rows and columns, is a versatile and widely employed data type. It finds application in numerous domains, including finance, healthcare with electronic health records, education, census data, and supply chain management. Its inherent structure makes it adaptable to a diverse range of data-driven tasks. 
% The generation of synthetic tabular data with differential privacy predominantly falls into two distinct categories: Statistics-based methods and deep learning-based methods. We investigate these methods in both centralized and decentralized settings. 

In this paper, we present a comprehensive review of existing differential private tabular data synthesis methods. The generation of differentially private synthetic tabular data primarily falls into two key categories: statistical methods and deep learning-based methods. We delve into both approaches, analyzing their strengths and limitations under both centralized and distributed settings. Furthermore, we offer insights into the unique challenges and considerations that arise in each context.

\begin{table}[h]
    \small
    \caption{Comparison with existing surveys}
    \begin{tabular}{c|c|c|c|c|c|c|c}
    \hline
    \hline
        \multirow{2}{*}{Paper}  &  \multirow{2}{*}{Year} &  \multirow{2}{*}{\tabincell{c}{Consideration of tabular \\data synthesis with DP}}  &  \multicolumn{2}{c|}{Centralized data synthesis} & \multirow{2}{*}{\tabincell{c}{Distributed \\data synthesis}} &  \multicolumn{2}{c}{Discussion on Evaluation} \\
        \cline{4-5} \cline{7-8}
        &&& S-M & DL-M &  & Fidelity/Utility & Privacy \\
        \hline
        \cite{bowen2020comparative} & 2020 & \checkmark  & \checkmark & & & & \\
        \hline
        \cite{bourou2021review} & 2021 &  & & \checkmark & & \checkmark & \\
        \hline 
        \cite{figueira2022survey} & 2022 &  & & \checkmark & & \checkmark & \\
        \hline
        \cite{ghatak2022survey} & 2022 & \checkmark  & \checkmark &\checkmark & & \checkmark & \\
        \hline
        \cite{xing2022non} & 2022 &  &\checkmark  & \checkmark & &\checkmark & \checkmark \\
        \hline 
        \cite{lu2023machine} &2023 &  & & \checkmark &  &\checkmark &\\
        \hline
        \cite{hassan2023deep} & 2023 & \checkmark  & & \checkmark & &  \checkmark & \\
        \hline
        \cite{10646785} & 2024&\checkmark & \checkmark & \checkmark & & &\\
        \hline

        \cite{bauer2024comprehensive} & 2024 &    & \checkmark & \checkmark & &   & \\
        \hline
        Ours & - & \checkmark & \checkmark & \checkmark & \checkmark & \checkmark & \checkmark \\
        \hline 
        \hline
    \end{tabular}

    \label{tab:survey} 
\end{table}

\noindent\textbf{Differences between this survey and others.} Currently, several synthesis surveys have been published, as shown in Table \ref{tab:survey}. Bourou et al. \cite{bourou2021review} reviewed several popular GAN-based models for tabular intrusion detection system data synthesis and experimentally evaluated their performance. However, this review exclusively focused on GAN-based models, without considering other methods. Figueira and Vaz et al. \cite{figueira2022survey} slightly extended the scope but still concentrated on GAN-based models for data synthesis. Xing et al. \cite{xing2022non} provided a broader review, covering data synthesis methods for non-imaging medical datasets, including both tabular and sequential data. Their discussion included statistical and deep learning-based methods, as well as evaluation metrics. Lu et al. \cite{lu2023machine} explored machine learning-based approaches for data synthesis and applications of synthetic data generation, addressing privacy and fairness concerns related to synthetic data. 

However, all the aforementioned papers discuss pure synthetic data generation methods, and none of them consider the methods with differential privacy protection. Bowen and Liu  \cite{bowen2020comparative} conducted an experimental study on differentially private data synthesis methods, focusing solely on statistical approaches. Hassan et al. \cite{hassan2023deep} explored the intersection of synthetic data and differential privacy, with a primary focus on deep generative models. Bauer et al. \cite{bauer2024comprehensive} conducted a comprehensive study of various model types suitable for synthetic data generation, including methods that incorporate differential privacy. Ghatak and Sakurai \cite{ghatak2022survey} considered the methods with differential privacy protection, but exclusively discussed data synthesis methods that emerged victorious in the NIST 2018 challenge. Hu et al. \cite{10646785} provided a review of differentially private data synthesis, including tabular data. However, their approach was more of a simple summary of existing methods rather than an in-depth analysis and discussion. 

Furthermore, all existing surveys focus on centralized data synthesis methods and do not address distributed data synthesis. In our paper, we target tabular data synthesis methods with differential privacy protection under both centralized and distributed settings. Additionally, we summarize and discuss various evaluation methods for the generated synthetic data, focusing on fidelity, utility, and privacy.

\vspace{2mm}
\noindent\textbf{Contributions of this survey.} This survey provides a comprehensive review of differential private data synthesis methods, focusing on tabular data. We consider two application scenarios: centralized data synthesis, where the data curator holds all users' datasets and aims to generate synthetic datasets for data analytics or sharing purposes, and distributed data synthesis, where data owners retain their data locally and collaborate with other parties for joint data synthesis. Our contributions are summarized as follows: 
\begin{itemize}
    \item We provide a thorough and comprehensive overview of existing methods for differentially private tabular data synthesis, along with the evaluation techniques used to assess their performance and effectiveness.
    \item Based on the synthetic data generation models, we categorize the primary approaches for data synthesis into two key research directions: statistical-based methods and deep learning-based methods, both applicable under two main scenarios: centralized and distributed data synthesis.
    \item We provide an in-depth review and analysis of existing methods for generating synthetic data, highlighting strengths and weaknesses in capturing attribute dependencies, modeling the distribution of attributes, computational complexity, and the noise scales introduced during the model learning process, etc. %the strengths, weaknesses, and challenges associated with each approach.
    \item By analyzing the state-of-the-art in the field, we discuss the research gaps and identify several promising future research directions to address the emerging challenges an advance the domain of private tabular data synthesis.
\end{itemize}

In this survey, we present the material in a tutorial manner, providing a clear introduction, comprehensive discussion, and valuable insights into the topics and methods. We aim to make the content accessible and informative for readers who are new to the subject as well as those looking to deepen their understanding. 

\vspace{2mm}
\noindent\textbf{Roadmap.} The rest of the paper is organized as follows: Section \ref{bk} provides background knowledge on tabular data synthesis and differential privacy. Section \ref{cdswdp} and Section \ref{ddswdp} discuss centralized data synthesis methods with differential privacy protection and distributed data synthesis methods with differential privacy protection, respectively. Section \ref{Sec:evaluation} introduces the synthetic data evaluation metrics. The research gaps and promising research directions are identified in Section \ref{frd}, and the survey is concluded in Section \ref{con}. 

\section{Background knowledge}\label{bk}

\subsection{Tabular data synthesis}

\subsubsection{Concepts.}This Section briefly presents the concept of tabular data synthesis.

\noindent\textbf{Tabular data.} Tabular data is data organized in a structured format with rows and columns, similar to a table. Each row represents a specific record or instance, such as a customer, transaction, or observation, and each column corresponds to a variable or attribute, such as name, age, or product price. 

\noindent\textbf{Data synthesis.} Data synthesis is the process of creating artificial datasets that replicate the structure, statistical properties, and relationships of real-world tabular data. The primary goal is to generate synthetic data that preserves the essential characteristics of the original data while ensuring privacy and enabling safe data sharing. This process is particularly valuable for scenarios where privacy concerns or data scarcity limit the use of real datasets.

\noindent\textbf{Centralized data synthesis.} Centralized data synthesis refers to the process of generating synthetic datasets where all the original data is collected, stored, and processed in a single, centralized location. A data curator or central authority typically holds the entire dataset and applies data synthesis techniques to produce a synthetic version. 

\noindent\textbf{Distributed data synthesis.} Distributed data synthesis involves generating synthetic datasets from data that is stored and processed across multiple locations or nodes. Each data owner retains control over their local data and collaborates with other parties to jointly create synthetic data without centralizing the original datasets. 

\subsubsection{Challenges.}Tabular data synthesis poses several challenges, particularly due to the complexity and diversity of tabular datasets compared to other data types like images or text. Some of the key challenges include:

\noindent\textbf{Data heterogeneity.} Tabular data often includes a mix of different data types (e.g., categorical, numerical and ordinal). Modelling the relationships between these different types of features is challenging. 

\noindent\textbf{Data distribution complexity.} First, many features in tabular data may not follow a simple distribution, and some may have multiple peaks (multi-modality), making it hard to model these distributions accurately. Additionally, real-world tabular data is often imbalanced, with skewed distributions in certain classes, it is challenging to represent the imbalance. 

\noindent\textbf{Feature dependencies.} The relationships between features can be highly complex (e.g., non-linear correlations). Capturing these relationships accurately in the synthetic data is challenging. 

Generating synthetic tabular data with differential privacy introduces additional challenges beyond those involved in general tabular data synthesis. These include the high sensitivity and dimensionality of the tabular data, which require significant noise to ensure privacy, and the cumulative noise can substantially degrade the overall quality of the synthetic data. Therefore,  It is challenging to balance the utility and privacy of the synthetic data. 
 
\subsubsection{Privacy disclosure in synthetic data.}One of the primary purposes of generating synthetic data is to enhance data privacy, facilitating data sharing without compromising users' sensitive information. However, research \cite{an2023comparison,trindade2024synthetic} have shown that sensitive information can still be disclosed if an attacker possesses some background knowledge about the victim. Several types of attacks can be performed on synthetic data, including the following: 

\noindent\textbf{Re-identification attack.} Re-identification \cite{jordon2021hide} attacks typically exploit auxiliary information or background knowledge that an attacker possesses about the individuals in the dataset. By correlating this additional information with the synthetic data, the attacker can identify specific individuals and extract sensitive information.

\noindent\textbf{Inference attack.} An inference attack occurs when an attacker deduces sensitive information about individuals from synthetic data by exploiting statistical properties, patterns, or background knowledge. This type of attack does not necessarily re-identify individuals but can still extract confidential information. It includes membership \cite{zhang2022membership} and attributes \cite{annamalai2024linear} inference and correlation exploitation. 

\subsection{Differential privacy}

\subsubsection{Definition of differential privacy}

Differential privacy, introduced by Dwork et al. \cite{dwork2006differential} in 2006, is a provable privacy concept. %It ensures that the inclusion, exclusion, or modification of an individual's record in a dataset does not significantly affect the overall statistical outcomes. The formal definition is shown as follows. 
It ensures that changing one person’s data does not have a big effect on the result, making it hard to tell if that person was included or not, all while still getting useful insights from the data. In differential privacy, $\epsilon$ plays a crucial role as a privacy parameter that controls the trade-off between privacy and accuracy. It sets the upper limit on how much the algorithm's output can differ when one individual's data is changed. A smaller 
$\epsilon$ value means stronger privacy protection and also tends to introduce more noise into the data, potentially reducing the accuracy of the results. In contrast, a larger $\epsilon$ provides more accurate results but weaker privacy guarantees.
% \begin{definition}[Differential privacy \cite{dwork2014algorithmic}] 
% A randomized algorithm $\mathcal{M}$ with domain $\mathbb{N}^{|\mathcal{X}|}$ is ($\epsilon, \delta$)-differential privacy if for all $\mathcal{S}\subseteq$ Range($\mathcal{M}$) and for all $x,y\in \mathbb{N}^{|\mathcal{X}|}$ such that $||x-y||_1 \leq 1$: 
% \[Pr[\mathcal{M}(x)\in \mathcal{S}]\leq exp(\epsilon)Pr[\mathcal{M}(y)\in \mathcal{S}] +\delta \]
% \end{definition}
($\epsilon, \delta$)-differential privacy is also known as approximate differential privacy, which is an extension of the standard differential privacy framework. It introduces an additional parameter, $\delta$, which allows for a small probability of a stronger privacy loss than $\epsilon$ would permit. The pure $\epsilon$-differential privacy is a special case of the approximate differential privacy when $\delta=0$.

\subsubsection{Differential privacy mechanisms} 

Differential privacy is achieved by introducing randomness into statistical computations. Commonly used mechanisms include the Laplace and Gaussian mechanisms for numerical data and the Exponential mechanism for categorical data. 

% \begin{definition}[The Laplace Mechanism] 
%     Given any function $f:\mathbb{N}^{|\mathcal{X}|}\rightarrow \mathbb{R}^k$, the Laplace mechanism is defined as:
%     \begin{align}
%         \mathcal{M}_L(x, f(\cdot),\epsilon) = f(x) + (\eta_1,\cdots, \eta_m)
%     \end{align}
%     where $\eta_i$ are i.i.d. random variables drawn from $\mathcal{L}(0,\Delta f/\epsilon)$. 
% \end{definition}
% The Laplace mechanism preserves $(\epsilon, 0)$-differential privacy.

% \begin{definition}[The Gaussian Mechanism] 
%     Given any function $f:\mathcal{D}\rightarrow \mathbb{R}$, the Gaussian mechanism is defined as:
%     \begin{align}
%         \mathcal{M}_G(x, f(\cdot),\epsilon) = f(x) + (\eta_1,\cdots, \eta_m)
%     \end{align}
%     where $\eta_i$ are i.i.d. random variables drawn from $\mathcal{N}(0,\Delta_2f^2\sigma^2)$. 
% \end{definition}
% The application of Gaussian mechanism to function $f$ with sensitivity $\Delta_2f$ satisfies $(\epsilon,\delta)$-differential privacy. 

% \begin{definition}[The Exponential Mechanism]
%     The exponential mechanism $\mathcal{M}_E(x,u,\mathcal{R})$ selects and outputs an element $r\in \mathcal{R}$ with probability proportional to $exp(\frac{\epsilon u(x,r)}{2\Delta u})$. 
% \end{definition}

% The exponential mechanism preserves $(\epsilon, 0)$-differential privacy.

\noindent\textbf{Laplace Mechanism.} The Laplace Mechanism is one of the most widely used methods in differential privacy. It protects privacy by adding noise to the output of a function, and this noise is drawn from the Laplace distribution. The scale of noise is determined by the sensitivity of the function, which measures how much the function’s output could change by altering a single individual’s data and the privacy parameter $\epsilon$. The Laplace mechanism preserves $\epsilon$-differential privacy.

\noindent\textbf{Gaussian Mechanism.} The Gaussian Mechanism works similarly to the Laplace Mechanism but adds noise drawn from the Gaussian distribution instead of the Laplace distribution. The noise magnitude is determined by both the sensitivity of the function and the privacy parameters, $\epsilon$ and $\delta$. The Gaussian mechanism satisfies $(\epsilon,\delta)$-differential privacy. 

\noindent\textbf{Exponential Mechanism.} Unlike the Laplace or Gaussian mechanisms, the Exponential Mechanism introduces noise to the selection process by assigning a probability to each possible output based on a utility function that measures how desirable each option is. The probability of selecting an output increases exponentially with its utility value, ensuring that the most useful outputs are more likely to be chosen, but with a privacy-preserving layer of randomness. The exponential mechanism preserves $\epsilon$-differential privacy.

\subsubsection{Composition properties} 

Differential privacy has several composition properties that facilitate the analysis of more complex privacy-preserving algorithms.

% \begin{theorem}[Parallel Composition \cite{dwork2014algorithmic}] \label{sc}
% Suppose that a method includes $m$ independent randomized functions $\mathcal{M} =\{\mathcal{M}_1, \mathcal{M}_2, ..., \mathcal{M}_m \}$, each $\mathcal{M}_i$ provides $\epsilon_i$-differential privacy guarantee. 
% If each $\mathcal{M}_i$ is performed on a disjointed record of the entire dataset, 
% then $\mathcal{M}$ is $\max\{\epsilon_1,\cdots,\epsilon_m\}$-differentially private. 
% \end{theorem}

\noindent\textbf{Parallel Composition}. The Parallel Composition Theorem \cite{dwork2014algorithmic} states that if a method includes $m$ independent randomized functions, each providing its own $\epsilon$-differential privacy guarantee, and if each function operates on a distinct portion of the dataset, then the overall privacy guarantee for the entire method is determined by the function with the highest $\epsilon$. In other words, the privacy level of the combined method is as strong as the function with the weakest privacy protection (the largest $\epsilon$).

%The basic sequential composition is shown as follows. 

% \begin{theorem}[Sequential Composition \cite{dwork2014algorithmic}] \label{sc}
% Suppose that a method includes $m$ independent randomized functions $\mathcal{M} =\{\mathcal{M}_1, \mathcal{M}_2, ..., \mathcal{M}_m \}$, each $\mathcal{M}_i$ provides $(\epsilon_i, \delta_i)$-differential privacy guarantee. If these functions are performed on the same dataset sequentially, then $\mathcal{M}$ is $(\sum_{i=1}^m \epsilon_i, \sum_{i=1}^m \delta_i)$-differentially private. 
% \end{theorem}

\noindent\textbf{Sequential Composition}. The Sequential Composition Theorem \cite{dwork2014algorithmic} states that when multiple differentially private mechanisms are applied sequentially to the same dataset, the overall privacy loss accumulates. Specifically, if $m$ independent functions each provide $\epsilon$-differential privacy guarantees and are applied to the same data, then the combined privacy guarantee is the sum of the individual privacy guarantees. This means that the total privacy loss increases with each additional query or operation on the same dataset, as the more queries are made, the more information about the dataset could potentially be revealed.

% The general sequential composition theorem considers the biggest value of the privacy loss. The bound is not tight. 
% To provide a tighter bound, Dwork et al. consider the expectation of the loss, which results in the advanced composition shown as follows. 

%The general sequential composition theorem accounts for the maximum privacy loss, but this bound is loose. To achieve a tighter bound, Dwork et al. focus on the expected privacy loss, leading to the advanced composition detailed below.

% \begin{theorem}[Advanced Composition \cite{dwork2014algorithmic}]
% For all $\epsilon, \delta, \delta' \geq 0$, the class of $(\epsilon,\delta)$-differentially private mechansims satisfies $(\epsilon', m\delta+\delta')$-differential privacy under $m$-fold adaptive compsoition for:
% \begin{equation}
%     \nonumber \epsilon' =\epsilon \sqrt{2m\log(1/\delta')} + m\epsilon(e^\epsilon-1)
% \end{equation}

% \end{theorem}

\noindent\textbf{Advanced Composition}. The Advanced Composition Theorem \cite{dwork2014algorithmic} provides a refined privacy guarantee when applying differentially private mechanisms multiple times. In the context of $m$-fold adaptive composition, where multiple queries are applied adaptively to the same dataset, the theorem shows that the privacy loss grows slower than the basic composition rule suggests. Specifically, for ($\epsilon, \delta$)-differentially private mechanisms, the total privacy guarantee becomes $(\epsilon', m\delta+\delta')$-differential privacy, where $\epsilon' =\epsilon \sqrt{2m\log(1/\delta')} + m\epsilon(e^\epsilon-1)$. This results in a tighter bound on the cumulative privacy loss, offering stronger privacy protection compared to the simple summing of the privacy parameters, especially when the number of queries $m$ is large.

\noindent\textbf{Moments Accountant \cite{abadi2016deep}}. The Moments Accountant method is a powerful technique used to track and control cumulative privacy loss when applying differential privacy mechanisms multiple times, particularly in scenarios like deep learning. It works by measuring the privacy loss at each step. The approach extends beyond considering just the expectation, using higher-order moments to bound the tail of the privacy loss variable. For each $\lambda$-th moment, the Moments Accountant provides a tighter bound on the cumulative privacy loss compared to traditional composition rules.

\subsubsection{Relaxations of differential privacy}

Strict differential privacy often requires adding significant noise to the data or query results, which can substantially degrade the utility of the data. Relaxations, such as ($\epsilon, \delta$)-differential privacy, allow for a controlled trade-off between privacy and utility. In addition to ($\epsilon, \delta$)-differential privacy, several relaxations of differential privacy have been proposed, offering greater flexibility in handling data complexities and providing tighter bounds on privacy loss. In this section, we present two commonly used definitions in the literature, Rényi differential privacy and zero-concentrated differential privacy.

% $(\epsilon,\delta)$-differential privacy offers smaller cumulative loss under composition. However, the application of an advanced composition leads to a wide selection of possibilities for $(\epsilon(\delta),\delta)$-differentially private guarantees, which implicity comes with an $\epsilon-\delta$ tradeoff. 
% Finding an optimal value may be non-trivial. To solve such a problem, Renyi differential privacy was proposed. 

\begin{definition}[$(\alpha,\epsilon)$-RDP \cite{mironov2017renyi}]
    A randomized mechansim $\mathcal{M}:\mathbb{D}\rightarrow \mathbb{R}$ is said to have $\epsilon$-Renyi differential privacy of order $\alpha$, or $(\alpha,\epsilon)$-RDP for short, if for any adjacent $D,D'\in \mathbb{D}$ it holds that 
    \begin{equation}
        \nonumber D_\alpha(\mathcal{M}(D)||\mathcal{M}(D'))\leq \epsilon,
    \end{equation}
\end{definition}
where $D_\alpha(\mathcal{M}(D)||\mathcal{M}(D'))$ is the Renyi Divergence of order $\alpha>1$, specifically, 
\begin{equation}
    D_\alpha(\mathcal{M}(D)||\mathcal{M}(D')) = \frac{1}{\alpha-1}\log E_{x\sim \mathcal{M}(D)}[(\frac{Pr[\mathcal{M}(D)=x]}{Pr[\mathcal{M}(D') =x]})^{\alpha -1}]
\end{equation}

RDP provides a flexible and refined method for measuring privacy loss using Rényi divergence. This approach offers a more accurate measure of cumulative privacy loss and tighter bounds when multiple queries are composed. By parameterizing privacy loss with Rényi divergence, RDP enables a smoother and more controlled tradeoff between privacy and utility, making it easier to achieve an optimal balance \cite{kairouz2015composition}. It simplifies the accumulation of privacy loss, reducing the complexity of maintaining privacy guarantees throughout multiple stages of data analysis. 

\begin{definition}[Zero-Concentrated Differential Privacy (zCDP) \cite{bun2016concentrated}]
A randomized mechanism $\mathcal{M}: \mathcal{X}^n\rightarrow \mathcal{Y}$ is $\rho$-zero-concentrated differentially private if, for all $x, x'\in \mathcal{X}^n$ differing on a single entry and all $\alpha \in (1, \infty)$, 
\[D_\alpha(M(x)||M(x'))\leq \rho \alpha \]

where $D_\alpha(M(x)||M(x'))$ is the $\alpha$- Rényi divergence between the distribution of $M(x)$ and $M(x')$. %\rho-zCDP$ is defined to be $(0,\rho)$-zCDP
\end{definition}

zCDP uses concentration inequalities to provide a more refined measure of privacy loss. This allows for tighter control over the distribution of privacy loss, leading to more efficient privacy guarantees. One of the major advantages of zCDP is its simpler and more efficient composition properties. zCDP allows for the straightforward addition of privacy loss terms when composing multiple queries.

% The $\alpha$ does not change, and the composition rule of RDP is straightforward, which is shown as follows. 

% \begin{theorem}[Composition] 
% Let $\mathcal{M}$ be a sequence of adaptive mechanisms $\mathcal{M}_1,\cdot,\mathcal{M}_m$, where $\mathcal{M}_i, i\in [m]$ guarantees $(\alpha, \epsilon_i)$-RDP, then $\mathcal{M}$ satifsifes $(\alpha, \sum_{i=1}^m\epsilon_i)$-RDP. 

% The definitaion of $\epsilon$-differential privacy conincides with $(\infty, \epsilon)$-RDP, where $(\infty, \epsilon)$-RDP implies $(\alpha, \epsilon)$-RDP for all finite $\alpha$.
% In turn, and $(\alpha, \epsilon)$-RDP implies $(\epsilon_\delta, \delta)$-differential privacy for any $\delta>0$. The following proposition shows the details. 

% \begin{proposition}[From RDP to ($\epsilon,\delta)$-DP]
% If $\mathcal{M}$ is an $(\alpha,\epsilon)$-RDP mechanism, 
% it also satisfies  $(\epsilon+\frac{log(1/\delta)}{\alpha -1}, \delta)$-differential privacy for any $0<\delta<1$. 
% \end{proposition}

% The moments' accountant and the RDP accountant are basically equivalent and lead to identical outcomes when translated to the $(\epsilon,\delta)$-DP language. 
% \end{theorem}

\section{Centralized data synthesis with DP} \label{cdswdp}

Centralized data synthesis involves generating synthetic data from a centralized server or database, where all original data is collected and stored in a single location. In this setup, the data curator is trusted and aims to release differentially private synthetic data that prevents sensitive information from being disclosed to third parties. %The methods for centralized data synthesis can be broadly categorized into two groups: statistical-based methods and deep learning-based method. 

\subsection{Statistical method}
Statistical methods emphasize maintaining the data distributions
by modelling the joint distribution of attributes and subsequently generating samples from
this model. It ensures that the synthetic data maintains identical statistical characteristics as those observed in the original dataset. %It has demonstrated exceptional performance when applied to tabular data. 
We categorize the methods into several groups based on the statistical models used to generate the synthetic data. Table. \ref{tab:statistics} summarizes the proposed methods. 

\subsubsection{Copula} \label{Sec:copula}

A copula is a statistical concept used to model and analyze the dependence structure between random variables. Copulas works by providing a mathematical framework to separate the modelling of the individual marginal distributions of random variables from the modelling of their joint dependence structure. Sklar's Theorem \cite{sklar1959fonctions}, formulated by Henry Sklar in 1959, is a fundamental theorem in copula theory. It states that any joint cumulative distribution function (CDF) of multiple random variables can be expressed in terms of their individual marginal CDFs and a copula function. Let $F(x_1), F(x_2), ..., F(x_n)$ be the marginal CDFs of $n$ random variables $x_1, x_2, ..., x_n$, and $C(u_1, u_2, ..., u_2)$ be a copula function of $n$ variables. Then, the joint CDF $F(x_1, x_2, ..., x_n)$ of these random variables can be represented as: 
\begin{equation}
    F(x_1, x_2, ..., x_n) = C(F(x_1), F(x_2), ..., F(x_n))
\end{equation}
In this equation, $C(u_1, u_2, ..., u_2)$ represents the joint distribution of the random variables with uniform marginal distributions. The general process of using copulas is to first transform your data into uniform marginal distributions, often through the use of CDFs of the individual variables. These uniform variables are then subjected to a copula function that captures how the variables' joint probabilities are related. Choosing a suitable copula function is a critical step. Copulas come in various families, including Gaussian, Clayton, Gumbel, and Frank, each tailored to different types of dependence structures. 

Selecting the appropriate copula family can be challenging.  Li et al. \cite{li2014differentially} and Asghar et al. \cite{asghar2020differentially} employed a Gaussian copula to model the joint distribution of the data, drawing inspiration from the common observation \cite{nelsen2007introduction} that many real-world high-dimensional datasets often exhibit Gaussian dependence structures. The main advantage of using a Gaussian copula lies in its efficiency, with a run-time that scales quadratically with the number of attributes. However, it assumes a linear relationship between variables \cite{xiao2019matching}. This can be problematic when modelling financial assets or other variables that do not have a truly Gaussian distribution or exhibit non-linear dependencies. To capture more complex dependence structures. Gambs et al. \cite{gambs2021growing} adopted vine copulas, which is a specific type of copula that decompose the multivariate functions of the popular into multiple `couples' of the copulas. The computational burden increases significantly in parallel. 

Estimating copula functions under differential privacy is a straightforward process that involves computing the DP marginals and the DP correlation matrix. Various well-established techniques \cite{acs2012differentially,xiao2010differential,xu2013differentially} are available for estimating DP marginal histograms, typically achieved by introducing Laplace or Gaussian noise to the histogram statistics. Since the introduction of Differential Privacy noise can be applied to individual attributes, confining its magnitude to the dimension of every single attribute, which often proves considerably smaller than when addressing joint distributions. However, when dealing with the DP correlation matrix, a more tailored approach is necessary due to the high global sensitivity exhibited by most correlation matrices, leading to excessive noise. 

\vspace{2mm}
\noindent\textit{Lessons learned and discussion.} Copula-based modelling is primarily designed for continuous variables, but existing works \cite{li2014differentially,asghar2020differentially,gambs2021growing} attempt to extend its applicability to discrete data. However, it is important to note that this extension typically applies to ordinal data with a sufficiently large domain. Even in these cases, the discrete data may need to be approximated or transformed into continuous data for accurate modelling.

\begin{table}[htbp]
  % \centering
   \small
    \caption{Statistical methods}
   % \vspace{0.2cm}
    % \vspace*{\fill} 
     %  \hspace*{5cm}
\begin{adjustbox}{angle=90,center}
    \begin{tabular}{|c|c|c|c|c|c|c|c|c|c|}
    \hline
    \multirow{2}{*}{Method}&\multirow{2}{*}{Reference} & \multicolumn{2}{c|}{\tabincell{c}{Marginal\\ selection}} & \multicolumn{2}{c|}{Noise addition} & \multicolumn{2}{|c|}{\tabincell{c}{Synthetic data \\generation}} &\multicolumn{2}{c|}{\tabincell{c}{Synthetic data\\ evaluation}} \\
    \cline{3-10} 
    & & \tabincell{c}{Selection \\method} & \tabincell{c}{Dependence \\evaluation} & \tabincell{c}{Privacy \\mechanism} & \tabincell{c}{Privacy \\accountant}  & Tool & method & \tabincell{c}{Utility\\ evaluation} & \tabincell{c}{Privacy \\evaluation} \\
    \hline
   \multirow{3}{*}{Copula} &   DPCopula \cite{li2014differentially}  & 1-m & GCF &  L& CP & Copula & Sampling & RQ & N \\
       \cline{2-10}
     &  COPULASHIRLEY \cite{gambs2021growing} &1-m & GCF&L &CP &Copula & Sampling & \tabincell{c}{KSD,MCD \\Classification \\Regression} & Y  \\
        \cline{2-10}
       &    DPSynthesizer  \cite{li2014dpsynthesizer} & 1-m & GCF & L & CP  & Copula & Sampling & RQ & N\\
    \hline
     \multirow{7}{*}{PGM} & PrivBayes \cite{zhang2017privbayes} & BN + IG & MI & L & CP & BN & Sampling & \tabincell{c}{k-m \\DWpre} & N \\
      \cline{2-10}
   & \tabincell{c}{PrivBayes \\improved \cite{bao2021synthetic}} & BN + IG & ID& AG & - & BN & Sampling & \tabincell{c}{Clustering \\Classification \\Regression } & N \\
      \cline{2-10}
   & FAPrivBayes \cite{ma2023improved} & BN + adaptive & MI & L & CP  & JT & Sampling & \tabincell{c}{k-m \\Classification} & N \\ 
      \cline{2-10}
   & PrivMRF \cite{cai2021data}  & MRF + IG & ID& AG & - & MRF + JT & Sampling & \tabincell{c}{k-m \\Classification} & N \\
      \cline{2-10}
   & Private-PGM \cite{mckenna2019graphical} &- & - & L & CP & MRF & Sampling & k-m & N \\
      \cline{2-10}
    &MST \cite{mckenna2021winning} & MST+IG & MI & G & RDP  & Private-PGM & Sampling & k-m, RQ & N \\
      \cline{2-10}
  &  JTree \cite{chen2015differentially} & SVT + Opt & MI & L & CP & JT & Sampling & \tabincell{c}{k-m \\Classification} & N \\
    \hline
  \multirow{8}{*}{Query} &   MWEM \cite{hardt2012simple} & QS & - & E+L & CP  & JD &Sampling & k-m, RQ & N \\
     \cline{2-10}
   & \tabincell{c}{DualQuery \cite{gaboardi2014dual} } & QS & - & E+L & CP & JD & Sampling & k-m & N \\
     \cline{2-10}
   & \tabincell{c}{DQRS \cite{vietri2020new} } & QS & - & E+L & CP & JD & Sampling & k-m & N \\
     \cline{2-10}
  &  FEM/sepFEM \cite{vietri2020new} & QS & - & E+L & CP & JD & Sampling & k-m & N\\
     \cline{2-10}
   & AIM \cite{mckenna2022aim} & QS + adaptive &-&E+G & zCDP  & Private-PGM & Sampling & k-m & N \\
   \cline{2-10}
   & PEP \cite{liu2021iterative} & QS & - & E+G & zCDP & JD & Sampling & k-m & N \\
     \cline{2-10}
   & RAP \cite{aydore2021differentially} & QS & - & G & zCDP & - & Opt + RR & k-m & N \\
   \cline{2-10}
   & RAP++ \cite{vietri2022private} & QS& - & G & zCDP & - & Opt & \tabincell{c}{k-m, LQ \\Regression} & N \\
      \cline{2-10}
   & PRIVATE-GSD \cite{liu2021iterative} & QS& - & E+G & zCDP & - & Opt & \tabincell{c}{k-m, LQ \\Regression} & N \\
    \hline
  \multirow{2}{*}{Others} &  DPPro \cite{xu2017dppro} & - & - &G &CP & - & Projection & \tabincell{c}{LQ \\Classification} & N \\
    \cline{2-10}
   & PrivSyn \cite{zhang2021privsyn} & Opt + Greedy& ID & G & zCDP  & - & GUM & \tabincell{c}{k-m,RQ\\Classification} & N \\
    \hline 
    \multicolumn{10}{c}{ \begin{tabular}[c]{@{}p{20cm}@{}}\footnotesize{1-m: 1-way marginal;BN: Bayesian Network; JD: Joint distribution; CP: composition property; L: Laplace mechanism; G: Gaussian mechanism; AG: Analytic Gaussian Mechanism \cite{balle2018improving}; E: Exponential mechanism; IG: information gain; QS: Query set; ME: Maximum entropy; DWpre: Dimension-wise prediction; k-m: k-way marginal; Opt: Optimization; zCDP:Zero-Concentrated differential privacy;GUM:Gradually update method; RQ: Range query; QS: Query set; GCF: Gaussian Copula Function; KSD:Kolmogorov-Smirnov distance;MCD: Mean Correlation Delta;SVT: Spare Vector Technology; RR: Randomized Rounding; LQ: Linear query }Identity attack\end{tabular}}
    \end{tabular}
  \end{adjustbox}  
 %   \vspace{0.2cm}

    \label{tab:statistics}
    
\end{table}

\subsubsection{Probabilistic graphical models}\label{Sec:pgm}

Probabilistic graphical models (PGMs) are a class of statistical models that use graphical representations to express and manipulate the joint probability distributions over a set of random variables. There are two main types of PGMs: Bayesian networks and Markov networks (also known as Markov random fields). Besides, Junction Trees play a crucial role in probabilistic inference within PGMs, primarily due to their efficiency in computing marginal probabilities, making them indispensable for handling complex probability distributions. Therefore, we categorize the methods into three main groups: methods based on Bayesian networks, approaches centered around Markov networks, and techniques leveraging Junction Trees. 

\vspace{2mm}
\noindent\textbf{Bayesian Network.} A Bayesian network utilizes a directed acyclic graph-based structure, where nodes serve as representations for random variables or events, and the directed edges convey probabilistic relationships among these variables \cite{stephenson2000introduction}. The joint distribution within the dataset can be estimated through the conditional distributions of attribute-parent pairs, as demonstrated below.
\begin{equation}\label{bne}
    Pr[X_1,X_2,\dots,X_d] = Pr[X_1] Pr[X_2|X_1]\dots Pr[X_d|X_1,\dots,X_{d-1}],
\end{equation}
where $X_i$ represents the attribute of the dataset and $d$ is the number of attributes. 

The procedure for constructing a Bayesian Network and estimating the joint distribution involves several key steps. Initially, the selection of attribute-parent pairs is essential. Once these pairs are identified, the subsequent step is to create a collection of conditional distributions for these selected attribute-parent pairs. To enhance privacy and confidentiality, it is crucial to incorporate differential privacy noise into all these computational steps. The synthetic data can then be sampled from the approximated distribution. 

\noindent\underline{\textit{Attribute-parent pair selection.}} 
The choice of attribute-parent pair plays a critical role in shaping the estimated joint distribution, aiming to approximate the dataset distribution closely. PrivBayes \cite{zhang2017privbayes} is the representative work. It employs the KL-divergence to assess the distance between two distributions. The smallest distance is achieved by maximizing the mutual information between the attribute denoted as $X_i$ and its corresponding parent set $\Phi$. To accomplish this, they employ a greedy approach to select attribute-parent pairs with maximal mutual information, thereby refining the model's fidelity to the original data distribution. 
The Exponential mechanism is employed to select the attribute-parent pair to achieve differential privacy in this process. In this context, the Exponential mechanism's score function is defined as the mutual information. It is worth emphasizing that mutual information exhibits high sensitivity, which results in the introduction of significant noise during the privacy-preserving operation \cite{zhang2017privbayes}. %Specifically, let $I$ represent the mutual information, and as demonstrated in \cite{zhang2017privbayes}, the sensitivity of mutual information $S(I)$ is shown to be greater than $log(n)/n$, where $n$ represents the number of samples. This sensitivity value is notably large when compared to the range of mutual information, which typically equals $1$ for binary distributions. 
To address this problem, in their subsequent research \cite{bao2021synthetic}, they introduce a solution based on utilizing a metric that quantifies the divergence between the joint distribution of attribute pairs, denoted as $P(A_i, A_j)$, and their product distribution, expressed as $P(A_i) \otimes P(A_j)$. It offers a sensitivity value of $2$, which is much smaller compared with its range $n$. 
In addition, Ma et al. \cite{ma2023improved} have introduced an approach to enhance the efficiency of network construction. Their method involves reducing the scale of candidate parent node sets. To achieve this, they assess the significance of each node by calculating the sum of its score function in relation to other nodes. Nodes with higher summed values exert greater influence on the network and are added earlier, up to a predefined threshold. 
This approach does come at the cost of consuming an additional privacy budget to perturb the importance of nodes. However, it proves effective in enhancing computational efficiency, particularly for datasets with a growing number of attributes when employing the exponential mechanism for attribute-parent pair selection.

\noindent\underline{\textit{Conditional distribution generation.}} 
The conditional distribution can be derived directly from the joint distribution. Once we have identified the attribute-parent pair, denoted as $X_i$ and $\Phi_i$, we can obtain the joint distribution, represented as $Pr[X_i, \Phi_i]$. A simple approach to introduce differential privacy is to inject Gaussian or Laplace noise into this distribution. However, Eq. \ref{bne} highlights the need for a set of high-dimensional marginal distributions. To mitigate dimensionality issues, Zhang et al. \cite{zhang2017privbayes} introduced the concept of a $k$-degree Bayesian Network, where each attribute is constrained to have at most $k$ parents. This reduces the scale of noise added to the dataset but may capture fewer correlations between attributes. It is important to acknowledge the practical trade-off involved in making this choice.

\noindent\textit{Synthetic data generation.} 
Once the conditional distribution is obtained, we have two options for generating synthetic data. First, we can directly sample synthetic data points from the distribution by computing the probabilities for each element within the variable's domain. Alternatively, we can opt for a sequential approach where attributes are sampled one by one in accordance with Eq. \ref{bne}, with each attribute being sampled based on its conditional distribution rather than the joint distribution. The sequential approach allows for a more efficient sampling process. To ensure the generation of tuples with low, yet non-zero frequencies, Bao et al. \cite{bao2021synthetic} introduced an alternative sequential approach. In this method, they convert the marginal distributions into histograms by scaling the probabilities based on the number of tuples to be generated. Subsequently, they create tuples one at a time, continuously adding to the histogram until they achieve the desired number of tuples. 

\vspace{2mm}
\noindent\textit{Lessons learned and discussion.} Bayesian networks have the flexibility to model both discrete and continuous variables, making them suitable for a wide range of data types. However, they heavily rely on assumptions about the conditional dependencies between variables \cite{krapu2023review}. If these assumptions are incorrect or incomplete, the synthetic data generated by the network may not accurately reflect the true data distribution. In addition, when estimating the joint distribution, there is a constraint placed on the number of marginals used, typically limited to a maximum of $d$ marginals, where $d$ represents the number of attributes. Consequently, capturing all important dependencies among the attributes becomes a challenging task. Furthermore, constructing Bayesian networks can be quite challenging, as it involves navigating a vast search space of potential network structures. Additionally, conducting inference in large Bayesian networks can be computationally expensive, posing challenges in efficiently generating synthetic data, especially for datasets with numerous variables. 

\vspace{2mm}
\noindent\textbf{Markov network.} 
Markov networks, also named Markov random fields, are one of the most widely used graphical models. In Markov networks, nodes represent random variables, while edges capture dependencies between them. Unlike Bayesian networks, Markov networks use undirected edges, making them well-suited for scenarios where relationships between variables are not easily represented by a causal hierarchy. The key feature of the Markov network is that it uses potential functions, denoted as $\phi_c(X_c) = e^{\theta(X_c)}$, to describe the relationships between nodes (random variables) in the cliques (a subset of nodes fully connected) of the graphical model. $\theta$ is the parameter corresponding to $X_c$. The joint probability distribution over all variables is defined as a product of these potential functions: 
\begin{equation}\label{eq:mn}
    Pr[X_1, X_2, \cdots, X_d] = \frac{1}{Z} \prod_{c\in \mathcal{C}}\phi_c(X_c)
\end{equation}
where $\mathcal{C}$ is the set of all cliques and $Z$ is the normalization factor.

Several key steps are involved in constructing a Markov network. Initially, a dependency graph is created to represent the interactions between variables. The noisy marginals are generated by adding Laplace or Gaussian noise to the marginal statistics. 
Parameters for the potential functions are then determined using statistical inference techniques, such as maximum likelihood estimation. The joint distribution is then estimated as described in Eq. \ref{eq:mn}. Finally, the synthetic data can be generated using the joint distribution information. 

\underline{\textit{Dependency graph construction.} } Constructing the dependency graph plays a crucial role in capturing attribute correlations. The general idea of constructing the dependency graph is to insert the edges to the graph when two attributes exhibit a high degree of correlation as measured by certain metrics, such as mutual information. Researchers often focus on modelling low-order correlations to mitigate the challenges posed by high-dimensional data and ensure differential privacy. Chen et al. \cite{chen2015differentially} assessed the mutual information between pairs of attributes and added an edge to the graph if the noisy mutual information is larger than a threshold. Besides, they utilize sampling techniques to amplify the privacy level to reduce the noise added to the mutual information calculation. A potential problem of the algorithm developed in \cite{chen2015differentially} is that the constructed dependency graph may result in quite sizable cliques. When this occurs, the marginal distribution of the clique becomes high-dimensional, demanding a significant amount of noise for privacy protection. Cai et al.\cite{cai2021data} proposed an iterative approach that greedily selects the node pairs with larger noisy scores, and inter the edges into the graph, while ensuring that the maximal clique in the triangulated graph remains below a specified threshold, which ensures there is no over-size clique in the tree.  McKenna et al. \cite{mckenna2019graphical} proposed a method, named Private-PGM, that aims to infer a data distribution by formulating an optimization problem that aligns the produced marginals closely with the observed ones. It does not offer a way to determine which marginals should be selected initially. One key consideration when working with Private-PGM is that the quality of the synthetic data it generates is heavily reliant on the accuracy of the provided marginals. To mitigate this issue, in their following work \cite{mckenna2021winning}, they constructed a maximum spanning tree to select the marginals combined with some selection rules, then employed Private-PGM as a post-processing tool for distribution estimation given noisy marginals.

\underline{\textit{Synthetic data generation.}} Once the noisy marginals are obtained and the parameters of the potential functions are estimated, the joint distribution can be determined using Eq. \ref{eq:mn}. Following this, synthetic data can be sampled based on this joint distribution. However, as the dataset's dimensionality increases, Markov networks can become complex, resulting in computationally expensive when sampling from high-dimensional joint distributions. Junction tree is commonly applied to ensure efficient and precise inference \cite{cai2021data}. 

\vspace{2mm}
\noindent\textit{Lessons learned and discussion.} Compared to Bayesian networks, Markov networks offer greater flexibility in modelling relationships between variables. However, learning the optimal structure of a Markov network from data can be computationally intensive and challenging, particularly as the number of variables increases. Estimating the parameters that define the potential functions, which describe how variables interact within the network, can also be difficult when data are scarce or the network structure is complex. Additionally, sampling from Markov networks, especially those with complex and densely connected structures, can be challenging and slow. Therefore, the junction tree algorithm is often utilized for probabilistic inference to manage these complexities effectively. 

\vspace{2mm}
\noindent\textbf{Junction tree.} \label{section:jt}
A junction tree is a data structure and associated algorithm used to facilitate efficient probabilistic inference in Markov networks. By constructing a junction tree, one can transform the Markov network into a tree-structured graphical model, which enables the calculation of joint and marginal distributions. 
Constructing a junction tree from a given dependency graph involves several steps. First, the graph must be triangulated by adding edges to ensure that any new edge between non-adjacent nodes forms a triangle, creating a chordal graph. Next, maximal cliques are identified within this triangulated graph; these are subsets of nodes that are fully connected and cannot be expanded without losing this connectivity. Then, for each pair of adjacent cliques, separator sets are determined. These sets consist of nodes that are common to both cliques and serve to separate them. Finally, the junction tree is constructed by representing each maximal clique as a node and each separator set as an edge connecting these nodes. 

% \begin{figure}
%     \centering
%     \includegraphics[scale=0.18]{Figures/Junctiontree.jpg}
%     \caption{An example of the creation of the junction tree (The figure is modified from \cite{chen2015differentially})}
%     \label{fig:jt}
% \end{figure}

\begin{figure}
    \centering
    \includegraphics[scale=0.18]{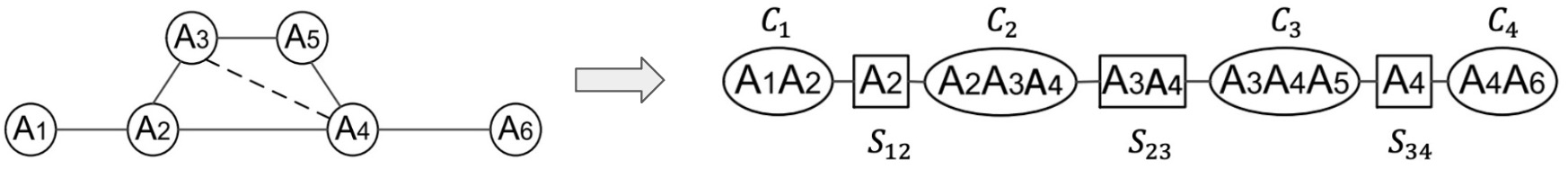}
    \caption{An example of the creation of the junction tree (The figure is modified from \cite{chen2015differentially})}
    \label{fig:jt}
\end{figure}

Figure \ref{fig:jt} illustrates an instance of a junction tree denoted as $T$ constructed from a dependency graph. When provided with an attribute set $A=\{A_1, A_2,\cdots, A_d\}$, the estimation of the joint distribution $Pr[A]$ is achieved by leveraging the marginals of a collection of cliques $C_i$ and their corresponding separators $S_{ij} = C_i \cap C_j$. 
The calculation \cite{chen2015differentially} is shown as follows. 
\begin{equation}\label{eq.jt}
    Pr[A] = \frac{\prod_{C_i\in T} Pr[C_i]}{\prod_{S_{ij}\in T}Pr[S_{ij}]}
\end{equation}

Once the junction tree is constructed, the marginal distributions of the cliques are determined. These are then used to estimate the overall joint distribution, enabling the generation of synthetic samples.

\noindent\underline{\textit{Noisy marginal generation.}} 
The straightforward approach to get a noisy margin is to compute the marginals of the cliques within the tree and the marginals of the separators. Then, add differential privacy noise to each of these marginals. Since the separators represent the shared attributes between adjacent cliques, we can always derive the marginal distribution of the separators from the cliques. Therefore, the focus should be on obtaining noisy marginals for the cliques, which we can then use to get the separator marginals and estimate the joint distribution as described in Eq.\ref{eq.jt}. 
The straightforward approach comes with a potential challenge. Specifically, when dealing with a junction tree containing a substantial number of cliques, a larger privacy budget is needed to obfuscate these marginals. This, in turn, can lead to a reduction in the accuracy of the estimated joint distribution. To mitigate this issue, similar strategies to those applied when deriving the separators have been considered.
In particular, Chen et al. \cite{chen2015differentially} proposed to merge the cliques to minimize statistical variance and obtain the noisy merged marginal distribution first. Subsequently, they derive the marginal distribution for each individual clique. 
%Secondly, the cliques may have a big size that includes many nodes that are fully connected. 

\noindent\underline{\textit{Synthetic dataset generation.}}
The joint distribution of the dataset, as defined by Eq. \ref{eq.jt}, presents computational challenges when it comes to sampling. To overcome this computational hurdle and efficiently sample data points from this distribution, it is rewritten as Eq. \ref{eq.jtchange} shown as follows. 
\begin{equation}\label{eq.jtchange}
    Pr[A] \approx Pr[C_1]\prod_{i=2}^{|C|}Pr[C_i \backslash S_{i}|S_{i}]
\end{equation}

Following Eq. \ref{eq.jtchange}, the synthetic dataset can be generated through a series of efficient local computations, as demonstrated by Cai et al. \cite{cai2021data} and previously by Chen et al. \cite{chen2015differentially}. %Specifically, we begin by selecting a clique, denoted as $\tilde{x}_{C_1}$, at random. For the selected clique $C_1$, we sample data records $\tilde{x}_{C_1}$ from its marginal distribution. We then proceed to generate data values for the remaining attributes in other cliques. We focus on cliques that share the same separator with the clique for which attributes have already been fully sampled. Specifically, for each such clique $C_i$, we examine the generated records $x_{S_{i}}$ for attributes in the separator $S_{i}$. We then sample data $\tilde{x}{C_i \backslash S{i}}$ from the conditional distribution $Pr[C_i \backslash S_i | S_i = \tilde{x}_{S_i}]$. We repeat the process until all attributes in the dataset have been sampled.

\vspace{2mm}
\noindent\textit{Lessons learned and discussion.} The junction tree algorithm is primarily associated with transforming Bayesian and Markov networks for efficient probabilistic inference. It is fundamentally a tool for simplifying and optimizing the inference process in complex graphical models. The advantage of using a junction tree is that it can directly sample from the joint distribution represented by the tree, using local computations within each clique and passing messages between cliques to maintain consistency across the entire network. But steps of triangulating the graph and identifying maximal cliques can be particularly challenging, especially for large or dense networks.

\subsubsection{Query-based method}

This stream of methods typically involves a series of queries that calculate the proportion of instances meeting certain criteria. It aims to respond to a wide range of statistical queries by creating a synthetic dataset from which the answers are derived.

%A stream of research portrays the synthetic data generation process as a zero-sum game \cite{hsu2013differential} involving two players: a data player and a query player. The data player's action set is the entire data universe $\chi$, and their strategy involves approximating the distribution of the true database. Conversely, the query player's action set is the query class $\mathcal{Q}$. For any given action pair $(x, q)$, where $x \in \chi$ and $q \in \mathcal{Q}$, the payoff is defined by $A(x,q) := q(D) - q(x)$, with $D$ representing the true dataset. In this game, the data player's objective is to minimize the payoff, whereas the query player aims to maximize it. 

Hardit et al. \cite{hardt2012simple} proposed a Multiplicative Weights Exponential Mechanism (MWEM) to approximate the distribution across the dataset's domain. It continually refines this approximation to enhance accuracy concerning both the private dataset and the desired query set by employing a combination of the Multiplicative Weights (MW) method \cite{hardt2010multiplicative} and the Exponential Mechanism. Specifically, it initializes a uniform distribution. Then, it identifies the query where the answer on the real data significantly differs from its corresponding answer on the approximate data, utilizing the Exponential Mechanism. %Then, it adjusts the weights of the approximating records positively contributing, scaling them up while simultaneously scaling down the weights of the records, causing a negative impact. 
It then increases the weights of the records that contribute positively to the approximation, while simultaneously decreasing the weights of records that have a negative impact.
%By the game interpretation, the MWEM algorithm employs the Multiplicative Weights approach \cite{hardt2010multiplicative}, enabling the data player to approximate the data distribution. Concurrently, the query player privately selects optimal queries as best responses using the Exponential mechanism. 

MWEM offers simplicity in implementation and usage, demonstrating commendable accuracy in practical applications \cite{hardt2012simple}. However, its efficacy diminishes when handling high-dimensional data due to its requirement to manipulate an object whose size scales linearly with the data universe's size. 
Several follow-up research tries to make improvements on top of MWEM. For example, Mckenna et al. \cite{mckenna2019graphical} scaled MWEM by replacing the multiplicative weights update step with a call to Private-PGM, which does not need to materialize the full contingency table. Later, they proposed AIM \cite{mckenna2022aim}, which refined MWEM-PGM \cite{mckenna2019graphical} by implementing a superior initialization strategy, allocating a fraction of the privacy budget to measure 1-way marginals. Additionally, they devised new selection criteria, such as imposing more significant penalties on larger marginals, when selecting queries. Furthermore, they introduced adaptive rounds and budget allocation adjustments, further enhancing the statistical accuracy of the method. Liu et al. \cite{liu2021iterative} improved the accuracy of MWEM by adaptively reusing past query measurements and selecting the synthetic data distribution with maximum entropy.

Gaboardi et al. \cite{gaboardi2014dual} model the synthetic data generation process as a zero-sum game, involving two players: a data player and a query player, which was first proposed by Hsu et al. \cite{hsu2013differential}. The data player's action set is the entire data universe $\chi$, and their strategy involves approximating the distribution of the true database. Conversely, the query player's action set is the query class $\mathcal{Q}$. Gaboardi et al. \cite{gaboardi2014dual} inverted the roles of the two players in the game, introducing a method named DualQuery. Specifically, it comprises a query player employing the no-regret learning algorithm and a data player that identifies optimal responses through the resolution of an optimization problem using MWEM. The proposed method is effective in managing high-dimensional data, but it doesn't lead to an improvement in accuracy. 
% Based on the observation that MW distribution changes slowly between rounds in the no-regret dynamics, Vietri et al. \cite{vietri2020new} reuse
% previously drawn queries to approximate the current MW distribution via rejection sampling, improving the accuracy of DualQuery. In addition, considering MW method maintains an entire distribution over the domain, which makes the MWEM runs in exponential time, they propose FEM and sepFEM that follow the same no-regret dynamics but importantly replace the MW method with two variants of the follow-the-perturbed-leader (FTPL) algorithm, both of which solve a perturbed optimization problem instead of maintaining an exponential-sized distribution. 
Vietri et al. \cite{vietri2020new} observed that the MW distribution changes slowly between rounds in no-regret dynamics. Leveraging this, they reuse previously drawn queries with rejection sampling to approximate the current MW distribution, enhancing the accuracy of DualQuery. Furthermore, since the MW method maintains a full distribution over the domain, causing MWEM to run in exponential time, they propose FEM and sepFEM. These methods, while still following no-regret dynamics, replace the MW approach with two variants of the follow-the-perturbed-leader (FTPL) algorithm, which solves a perturbed optimization problem instead of managing an exponentially large distribution.

To create synthetic data that most closely aligns with a noisy set of query measurements obtained from the original data. Several works use gradient-based optimization to solve this problem. Aydore et al. \cite{aydore2021differentially} proposed a relaxed adaptive projection mechanism (RAP), which adaptively uses a continuous relaxation of the Projection Mechanism \cite{nikolov2013geometry} to generate the synthetic dataset. It iteratively updates an initially synthesized dataset to minimize the $l_2$ distance between the noisy query results on the real dataset and those on the synthetic dataset. To conserve the privacy budget, rather than answering all the queries upfront, they start by answering a small number of queries. These are then projected onto a vector of answers consistent with a relaxed synthetic dataset. In subsequent rounds, they privately identify additional queries where the synthetic dataset underperforms, respond to these queries, and generate a new dataset through continuous projection. 
Vietri et al. \cite{vietri2022private} extended this research by introducing RAP++, which employs random linear projection queries to manage mixed-type data, eliminating the need for discretization of numerical features. Due to the differentiability requirement of the gradient-based optimization, they circumvent discretization by approximating non-differentiable queries with differentiable surrogate functions, which may introduce extra error due to the relaxation. To solve this problem, Liu et al. \cite{liu2021iterative} proposed using a generative algorithm, which does not require differentiability in the optimization objective.
In these methods, each iterative step involves updating the entire synthetic dataset to minimize the discrepancy between the real and synthetic data based on the new query responses. This process can be computationally demanding, especially as the size of the dataset and the number of queries increase. 

\vspace{2mm}
\noindent\textit{Lessons learned and discussion.} This group of methods typically requires a set of queries. Therefore, no need to make the selection of the marginals to estimate the joint distribution, all marginals are determined by the workload. An advantage of this approach is that it does not waste the privacy budget on selection processes, allowing the generated synthetic dataset to achieve more accurate performance to the queries in the set. However, if the set of queries does not comprehensively cover all relevant aspects of the data, the synthetic dataset might not adequately represent the characteristics of the original dataset beyond those queries.

\subsubsection{Other methods} 
In addition to the mainstream statistical methods we have discussed, several other approaches have been proposed in the literature. %This section will explore these alternative methods.

\vspace{2mm}
\noindent\textbf{Maximum entropy optimization.} Maximum entropy is a powerful and widely used principle for estimating joint probability distributions of multiple random variables, which has been widely adopted in the research of $k$-way marginal estimation \cite{qardaji2014priview, zhang2018calm}, where $k$ can be as large as the number of attributes. Privacy is ensured by introducing noise into the selected lower-dimensional marginals. The key idea behind maximum entropy is to find the most unbiased or least informative probability distribution that is consistent with the available information (e.g., the marginals) or constraints (e.g., the consistent or non-negative constraint). It seeks to maximize the entropy of the distribution while satisfying these constraints. Existing studies \cite{cormode2018marginal,tang2022marginal,qardaji2014priview} %on estimating joint distributions primarily focus on low-dimensional marginals. Using such a method to estimate the joint distribution for high-dimensional data would be computationally intensive. 
primarily focus on estimating low-dimensional marginals. Applying these methods to high-dimensional data would be computationally intensive.

\vspace{2mm}
\noindent\textbf{Projection-based method.} Xu et al. \cite{xu2017dppro} proposed the use of the Johnson-Lindenstrauss transformation to project high-dimensional datasets into a lower-dimensional space. According to Johnson-Lindenstrauss's theory, this transformation yields a reduced representation that preserves pairwise distances between points. Privacy is enhanced by adding noise to the projected dataset. While this method maintains the Euclidean distances between high-dimensional vectors, the resulting dataset often differs in shape from the original dataset, which may not be desirable. 

\vspace{2mm}
\noindent\textbf{Gradually update method.} Zhang et al. \cite{zhang2021privsyn} presented an alternative method in their work called the  ``gradually update method (GUM)" to generate the synthetic dataset utilizing the selected noisy marginal. This method begins with the initialization of a random dataset and then proceeds to iteratively update its records to ensure consistency with the provided marginals. The resulting data records generated using this method tend to generally align more closely to the noisy marginal statistics than those generated by methods like probabilistic graphical models. However, it is worth noting that the dataset updating process can encounter convergence issues that achieving convergence may not always be straightforward, making it a challenging aspect of this approach.

\subsection{Deep learning-based method}

Deep learning (DL) methods are widely utilized for image synthesis, typically handling homogeneous numerical data. However, there is an increasing interest in applying these techniques to tabular data, with adaptations being made to suit this purpose.
Table \ref{tab:dp-bm} summarizes the proposed methods, each of which is discussed in the following sections.

\begin{table}[!t]
    \centering
        \caption{Deep learning - based data synthetic methods}
        \vspace{0.2cm}
    \small
    \begin{tabular}{c|c|c|c|c|c|c}
        \hline
         \hline
        DL model & Methods &  \tabincell{c}{Privacy\\ mechanism} &  \tabincell{c}{Constant \\clipping}&\tabincell{c}{Privacy\\ accountant} & Utility evaluation & \tabincell{c}{Privacy \\ evaluation}  \\
        \hline
        \multirow{2}{*}{AE}&DP-AuGM\cite{chen2018differentially} & DPSGD & Y &MA  & classification & Y\\
        \cline{2-7}&
        DP-SYN \cite{abay2019privacy} & \tabincell{c}{DPSGD\\ DP-EM}& Y&MA  & \tabincell{c}{k-way marginal\\ classification\\ agreement rate} & N   \\
        \hline
        \multirow{3}{*}{VAE}&DP-VaeGM \cite{chen2018differentially} & DPSGD & Y & MA & classification & Y  \\
        \cline{2-7}&
        DPGM \cite{acs2018differentially} & \tabincell{c}{DP-k-means\\ DPSGD}& N & MA & linear query & N  \\
        \cline{2-7}&
        P3GM \cite{takagi2021p3gm} & \tabincell{c}{DP-PCA\\ DP-EM\\ DPSGD} &Y & RDP & \tabincell{c}{classification\\ k-way marginal} & N  \\ 
        \hline        
        \multirow{5}{*}{GAN}&DPGAN\cite{xie2018differentially} & DPSGD  & Y&MA & DWpre, DWpro &N \\
        \cline{2-7}& 
        dp-GAN \cite{frigerio2019differentially} &DPSGD & N&MA &DWpre&N  \\
        \cline{2-7}&
        DPNet \cite{fan2021dpnet} & DPSGD  &N& MA & \tabincell{c}{realism\\ k-way marginal} & N  \\
        \cline{2-7}  &      
        PATE-GAN \cite{jordon2018pate} & PATE  &-& MA  & classification & N  \\
        \cline{2-7}&
        G-PATE \cite{long2021g} & PATE &-& RDP & classification & N \\
        \hline
        \multirow{2}{*}{AE+GAN}&RDP-CGAN \cite{torfi2022differentially} &DPSGD& Y& RDP & MMD, DWpre & N  \\
        \cline{2-7}&        
        DP-auto-GAN \cite{tantipongpipat2021differentially} & DPSGD & Y& RDP& \tabincell{c}{DWpre\\ DWpro\\ k-way marginal} & N  \\
        \hline
        \multirow{3}{*}{GN}&GEM \cite{liu2021iterative} &\tabincell{c}{Exponential\\ Gaussian} & Y & zCDP & k-way marginal & N\\
        \cline{2-7}&        
        DP-MERF \cite{harder2021dp}  & Gaussian & Y & RDP & classification & N   \\
        \cline{2-7} &
        DP-HP \cite{vinaroz2022hermite} & Gaussian & Y & RDP & \tabincell{c}{k-way marginal\\ classification} & N  \\
        \hline
         \hline
    \end{tabular}
          \vspace{0.2cm}
          
    \footnotesize{MA:moment accountant; RDP: Renyi differential privacy; DWpre: Dimension-wise prediction; DWpro: Dimension-wise probability; GN: Generative Network}

    \label{tab:dp-bm}

\end{table}

\subsubsection{Autoencoder (AE)} The autoencoder is a highly prevalent unsupervised learning model with the primary objective of acquiring a compact data representation, often employed for dimensionality reduction \cite{cortes2024autoencoder,berahmand2024autoencoders}. This neural network architecture operates by simultaneously training an encoder, responsible for converting high-dimensional data points into lower-dimensional representations, and a decoder, tasked with reconstructing high-dimensional data from the compressed representation. This process allows the model to capture essential features within the data while minimizing the overall data volume. 

Chen et al. \cite{chen2018differentially} trained a differentially private autoencoder and made the encoder available for generating synthetic data by inputting the user's own data. The resulting synthetic data is then employed in downstream prediction tasks. However, the generated synthetic dataset is a low-dimensional data representation that differs from the original dataset in terms of data format. 
Abay et al. \cite{abay2019privacy} applied the expectation maximization function to optimize the output of the encoded data and generate the synthetic data by decoding the encoded data. 

\vspace{2mm}
\noindent\textit{Lessons learned and discussion.} Such group methods cannot produce arbitrary synthetic datasets. Typically, autoencoders are used in conjunction with other generative models, such as GANs, serving as a preprocessing step to prepare the input data for the generative model.

\subsubsection{Variational Autoencoder (VAE)} 
A Variational Autoencoder is a generative model that combines the principles of autoencoders and probabilistic modelling. Different from AE, which only tries to reduce the dimension, VAEs are designed to learn and represent complex, high-dimensional data in a lower-dimensional space while simultaneously capturing the underlying probability distribution of the data \cite{pinheiro2021variational}.  An encoder network in VAE maps the input data to a probability distribution in the latent space, while a decoder network generates data samples from this distribution. VAEs use variational inference to model uncertainty in the latent space, which allows for the generation of not just deterministic reconstructions but also diverse and expressive data samples. 

Chen et al. \cite{chen2018differentially} employed multiple private VAEs, with each VAE dedicated to generating synthetic data for a specific class. Their empirical findings revealed that training $n$ generative models achieves higher utility compared to training a single model in terms of prediction accuracy. Similarly, Acs et al. \cite{acs2018differentially}
divided the data using k-means clustering and subsequently trained separate VAEs for each partition. In line with the conclusion in \cite{chen2018differentially}, they find that using multiple models resulted in more accurate synthetic samples, as it prevented the mixture model from generating unrealistic synthetic data that could emerge from improbable combinations of very distinct clusters. Despite these methods offering advantages, the synthetic data might not follow the original distribution. This is because the synthetic data is generated independently across separate datasets. Rather than training multiple VAEs, Takagi et al. \cite{takagi2021p3gm} expanded the VAE architecture by employing a dimensional reduction function in lieu of embedding. Additionally, they fixed the encoder's mean to a constant value, thereby narrowing the search space and significantly accelerating the convergence speed, which in turn improves the model accuracy with a fixed privacy budget. 

\vspace{2mm}
\noindent\textit{Lessons learned and discussion.} In a VAE, a common assumption involves employing a Gaussian distribution in the latent space owing to its simplicity and ease of use. However, datasets with notable skewness or non-normal distributions may not conform well to this assumption. While it is possible to adapt the distribution of the latent space, the challenge lies in determining a suitable distribution that aligns with the specific characteristics of the data. In addition, more complex distributions might capture fine-grained details but could also increase model complexity and computational requirements. Furthermore, VAEs might struggle with imbalanced datasets, potentially leading to difficulties in generating representative samples for minority classes or rare instances.

\subsubsection{Generative Adversarial Network (GAN)}
Generative Adversarial Networks \cite{goodfellow2020generative} are a class of deep learning models consisting of two neural networks, a generator and a discriminator, engaged in a competitive game. The generator creates synthetic data samples from random noise or other input sources. The discriminator evaluates both the synthetic data generated by the generator and the real data samples to differentiate them. The parameters of the generator and discriminator are updated based on the computed loss to improve the performance. The training process continues until the generator achieves the ability to generate synthetic data that closely resembles real data, and the discriminator reaches a point where its accuracy in distinguishing between the two levels is off. Due to the mode complexity of GAN, the training samples are easily remembered by the model \cite{xie2018differentially}. To ensure privacy, two privacy models are considered in the literature, Differentially Private Stochastic Gradient Descent (DPSGD) \cite{abadi2016deep} and Private Aggregation of Teacher Ensembles (PATE) \cite{papernot2017semi}. 

\vspace{2mm}
\noindent\textbf{DPSGD.} The widely used framework DPSGD has been applied to the GAN training process, specifically adding noise to the gradient of the discriminator during training to provide provable privacy protection. 
Frigerio\cite{frigerio2019differentially}  and Fan et al. \cite{fan2021dpnet} optimized this process by reducing the clipping bound for each iteration, in turn, reduces the introduced noise. Fan et al. \cite{fan2021dpnet} further improved the performance by privately selecting the best model across all training epochs. Besides, they train an embedding model to capture the relationships between features. However, to save the privacy budget, the embedding mode is required to train on a public dataset.

\vspace{2mm}
\noindent\textbf{PATE.} PATE employs an ensemble of teachers trained on different subsets of data, ensuring that no single model has access to the entire sensitive dataset \cite{liu2023privacy}. 
In a typical GAN framework, there is a single discriminator trained in direct opposition to the generator. PATE-GAN \cite{jordon2018pate}, however, introduces $k$ teacher discriminators alongside a student discriminator. To ensure differential privacy, the student discriminator is only trained on records generated by the generator and labelled by the teacher discriminators. This framework limits the influence of individual samples on the model, providing strong differential privacy guarantees. However, the approach assumes that the generator can cover the entire real data space during training. If most synthetic records are labelled as fake, the student discriminator could be biased and fail to learn the true data distribution. Different from PATE-GAN, Long et al. \cite{long2021g} proposed an ensemble of teacher discriminators to replace the GAN's single discriminator. A differentially private gradient aggregator is incorporated to collect information from these teacher discriminators, which guides the student generator to improve synthetic sample quality. Instead of ensuring differential privacy for the discriminator, noise is added to the flow of information from the teacher discriminators to the student generator.

\vspace{2mm}
In addition, GANs can suffer from mode collapse \cite{zhang2018convergence}, where they generate limited varieties of samples, especially in complex and high-dimensional data spaces like tabular datasets. This can result in a lack of diversity in generated samples, failing to represent the full complexity of the original data distribution \cite{takagi2021p3gm}. Addressing these challenges often involves modifications to the GAN architecture and exploring novel training methods tailored for the tabula data generation task. Some efforts have been made in the literature, for example, Fan et al. \cite{fan2021dpnet} learned an embedding during the training process to capture the relationships between attributes using a public dataset. 
Conditional GAN \cite{mirza2014conditional} was applied to deal with the imbalanced label distribution. Specifically, it encodes the label as a condition vector to guide the generator to generate samples with the label. 
Long et al. \cite{long2021g} proposed to utilize a small privacy budget to estimate the class distribution in the training dataset and use the trained differentially private generator to generate data following the estimated class distribution. 
To overcome the model collapse problem of GAN, a variant of GAN, Wasserstein GAN \cite{arjovsky2017wasserstein}, often abbreviated as WGAN, was adopted \cite{uclanesl_dp_wgan,liu2024tabular}. It introduces the Wasserstein distance (also known as the Earth Mover's Distance) as a more stable and informative metric for training generative models. %It is worth noting that DPWGAN \cite{uclanesl_dp_wgan} achieved 5th place in the NIST DP Synthetic Data Challenge. 
Liu et al. \cite{liu2024tabular} proposed to simplify the neural network architectures of the discriminator to limit the capacity of the discriminator and then avoid the chance of gradient disappearance of the generator. 
Xu et al. \cite{xu2019modeling} proposed a mode-specific normalization based on the Gaussian Mixture Model to capture complex data distributions. However, this method does not offer privacy-preserving features.

\vspace{2mm}
\noindent\textit{Lessons learned and discussion.} GANs have achieved significant success in image generation \cite{xu2019ganobfuscator,chen2020gs,de2023review,liu2024precise}, which can be represented in a continuous space. However, their application to tabular data is still in the early stages. Unlike image data, tabular data possesses unique characteristics that present challenges for their adoption in this context. 
First, tabular data often includes mixed data types; second, the attributes of tabular data are usually correlated. However, the majority of work on GANs focuses on making the synthetic data visually resemble real data \cite{shmelkov2018good,tran2021data}, often overlooking these correlations between features. Third, the datasets may exhibit highly imbalanced data distributions. Without careful consideration, this can result in insufficient training for records with minority labels. 

\subsubsection{AE+GAN} One challenge with GANs is that they are primarily suited for continuous data types, whereas tabular data often includes a mix of both categorical and numerical data types. Autoencoders can effectively address this issue as they are capable of encoding categorical data into a numerical format using techniques such as one-hot encoding and label encoding. Additionally, embedding layers in autoencoders can transform sparse categorical variables into dense, lower-dimensional representations, making them more amenable to processing by neural networks. Therefore, the AE is used in conjunction with GANs to handle mixed data types in tabular datasets \cite{tantipongpipat2021differentially}. Besides, autoencoders are designed to compress data into a lower-dimensional latent space while still preserving the key characteristics of the original data, which aids in capturing the correlations between features. Torfi et al.\cite{torfi2022differentially} have enhanced this capability by integrating one-dimensional convolutional neural networks (CNNs) within autoencoders. 

\vspace{2mm}
\noindent\textit{Lessons learned and discussion.} The combination of AE and GAN allows for better data representation and generation, leveraging the strengths of both models. However,  Integrating multiple models might complicate the implementation of differential privacy mechanisms, making it challenging to ensure both privacy and utility.

\subsubsection{Generative Network (GN)} One stream of research proposes using a generator exclusively to create synthetic datasets. It employs techniques to quantify the distances between the information obtained from real data and synthetic data. The generator is designed to minimize this distance, thereby producing data that closely resembles the real data. 

Harder et al. \cite{harder2021dp} proposed using Maximum Mean Discrepancy (MMD) \cite{gretton2012kernel} to quantify the distance between distributions in a Hilbert space \cite{berlinet2011reproducing}. The distributions of real and synthetic datasets are transformed into the Hilbert space using kernel mean embeddings (KME), which compute the mean of kernel evaluations. Gaussian noise is added to the kernel mean embeddings of the real dataset to ensure privacy. Due to the resource-intensive nature of computing KMEs, they approximate the KMEs using the inner product of feature vectors. However, since a large number of random features are required to achieve a good approximation, the noise increases significantly. To address this issue, Vinaroz et al. \cite{vinaroz2022hermite} used Hermite polynomial features, capturing more information with fewer features. Additionally, to tackle the imbalance issue in tabular datasets, they propose allocating part of the privacy budget to obtain the statistics of class counts and modify the released mean embeddings by appropriately weighting the embedding for each class. Liu et al. \cite{liu2021iterative} defined the distance based on a set of queries. Specifically, they defined the loss function as the distance between the query results on the synthetic data distribution and the noisy query results on the real dataset. The work focuses on query release, optimizing the distribution to provide accurate query answers. The sampled synthetic data may not accurately represent the original data characteristics. 

\vspace{2mm}
\noindent\textit{Lessons learned and discussion.} Using a generative network with only a generator to produce synthetic data has several advantages. The architecture is simpler, avoiding the complexities and potential instability issues present in more complex generative models like GANs, and it makes privacy noise addition more straightforward. However, designing an effective loss function that accurately captures the data distribution differences can be challenging due to the limited feedback.

\section{Distributed data synthesis with DP}\label{ddswdp}

Decentralized data synthesis involves the collaboration and synthesis of data from multiple parties or sources. %Collaborative synthesis mitigates biases present in individual datasets and increases the statistical power of analyses, leading to more reliable outcomes. Collaborative data synthesis 
It becomes especially valuable when one party lacks a sufficient dataset for meaningful analysis or insights. 
In a decentralized setup, a semi-trust server typically coordinates the learning process. Rather than sending the raw dataset to the server, each client shares statistical or intermediate results from their local models. Together, they collaboratively learn a synthetic data generative model. To safeguard clients' data, differential privacy noise is incorporated into the information transmitted to the server. 
We categorize the existing works into two types: vertical data synthesis and horizontal data synthesis. Table \ref{tab:distributed_ds} summarizes these works. 

\begin{table}[h]
    \centering
        \caption{Distributed data synthesis with differential privacy}
    \small
    \begin{tabular}{|c|c|c|c|c|c|c|l|}
    \hline
        \multirow{2}{*}{\tabincell{c}{Data synthesis\\ method}} & \multirow{2}{*}{\tabincell{c}{Data \\partition}} & \multicolumn{2}{c|}{Model} & \multicolumn{2}{c|}{Architecture} &\multirow{2}{*}{Privacy}& \multirow{2}{*}{Key limitation} \\
        \cline{3-6}
        &  & Statistics & DL & MD & FL & &\\
        \hline
        DistDiffGen\cite{mohammed2013secure} & V & CT & & & &$\epsilon$-DP&Only suitable for two-party scenario \\
        \hline
        DPLT\cite{tang2019differentially} & V & BN & & & &$\epsilon$-DP&Limited to discrete attributes \\
        \hline
        VertiGAN\cite{jiang2023distributed} & V & & GAN &  \checkmark & &$(\alpha,\epsilon(\alpha))$-RDP & Cannot deal with imbalanced data \\ 
        \hline
        GTV\cite{zhao2023gtv} & V & & GAN & & \checkmark &-& Server cannot collude with any client \\
        \hline
        DP-SUBN \cite{su2016differentially} & H & BN  & & & & $\epsilon$-DP & Split privacy budget needs too much\\
        \hline
        Fed-TGAN\cite{zhao2021fed} & H & & GAN & & \checkmark & - & Joint distribution is not considered\\
        \hline
        DP-CTGAN \cite{fang2022dp} & H & & GAN & & \checkmark & $(\epsilon,\delta)$-DP & No privacy for the feature statistics \\
        \hline
        HT-Fed-GAN \cite{duan2022ht} & H & & GAN & & \checkmark & $(\epsilon,\delta)$-DP & High communication cost \\
        \hline 
        ATLAS \cite{wang2023atlas} & H & & GAN & \checkmark & & $(\alpha,\epsilon(\alpha))$-RDP & Cannot deal with the skewed dataset \\
        \hline
     %   DPD-fVAE\cite{pfitzner2022dpd} & H & & VAE & 
     %   SGDE \cite{lomurno2022sgde} & H & & VAE & 
        \hline
    \end{tabular}
                    \vspace{0.2cm}
                    
    \footnotesize{V: Vertical; H: Horizontal; CT: Contingency table; BN: Bayesian Network; MD: Multi-discriminator structure; FL: Federated learning structure}
    \label{tab:distributed_ds}
\end{table}

\subsection{Vertical data synthesis} 
Vertically partitioned data refers to scenarios where different subsets of attributes of the same dataset are stored across different parties or locations. It is quite common in various fields, including healthcare, finance, retail and more. 
Many organizations naturally organize their data based on different departments, divisions, or functional areas. For example, in a healthcare scenario where patient information is vertically distributed across various departments within a hospital. %The cardiology department holds data related to heart conditions, the pathology department stores information about lab test results, and the patient records department maintains demographic details, medical history, treatments, and prescriptions. 
Merging these segregated datasets helps to generate a comprehensive patient profile. In another scenario, datasets may originate from distinct organizations, like a bank with customer income records and an e-commerce company with customer purchase histories. By combining these disparate datasets, each organization gains valuable insights that augment their respective analyses by leveraging diverse information across sectors. Specifically, within the context of vertical data synthesis, each client holds a local dataset containing distinct features related to the same individuals. A specific challenge in vertical data synthesis is capturing the relationships between columns across datasets from different clients. 

\vspace{2mm}
\noindent\textbf{Statistical method.}
Mohammed et al. \cite{mohammed2013secure} introduced the initial method for releasing vertically partitioned data that forms an integrated data table. Their approach involves using a predefined attribute taxonomy tree as publicly available information, along with the distributed exponential mechanism to produce the generalized data table. Additionally, they apply Laplace noise to the true count at leaf nodes to guarantee differential privacy. However, this method is only suitable for a two-party scenario due to the limitations of various underlying cryptographic primitives. And the data utility deteriorates quickly with the increase in the number of
attributes. Tang et al. \cite{tang2019differentially} expanded the algorithm's capacity to handle a larger number of attributes by leveraging a latent tree model \cite{zhang2004hierarchical}. They utilized this model to capture the dependencies among attributes, followed by privatizing these latent tree parameters using a distributed Laplace protocol. The acquired model was then employed to generate synthetic data. The proposed method enhances data utility to a certain extent, yet the rise in data dimensionality could still result in significant utility loss. Moreover, its applicability remains restricted to discrete attributes. 

\begin{figure*}[ht]
     \centering
\subfigure[Multi-Discriminator structure]{
\includegraphics[scale=0.15]{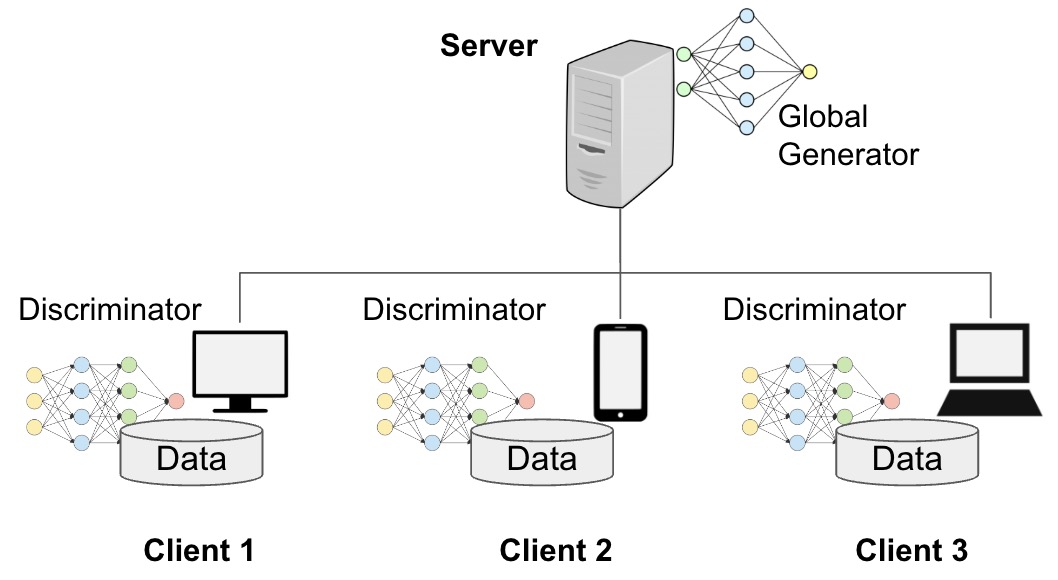}
}
\subfigure[Federated Learning structure]{
\includegraphics[scale=0.15]{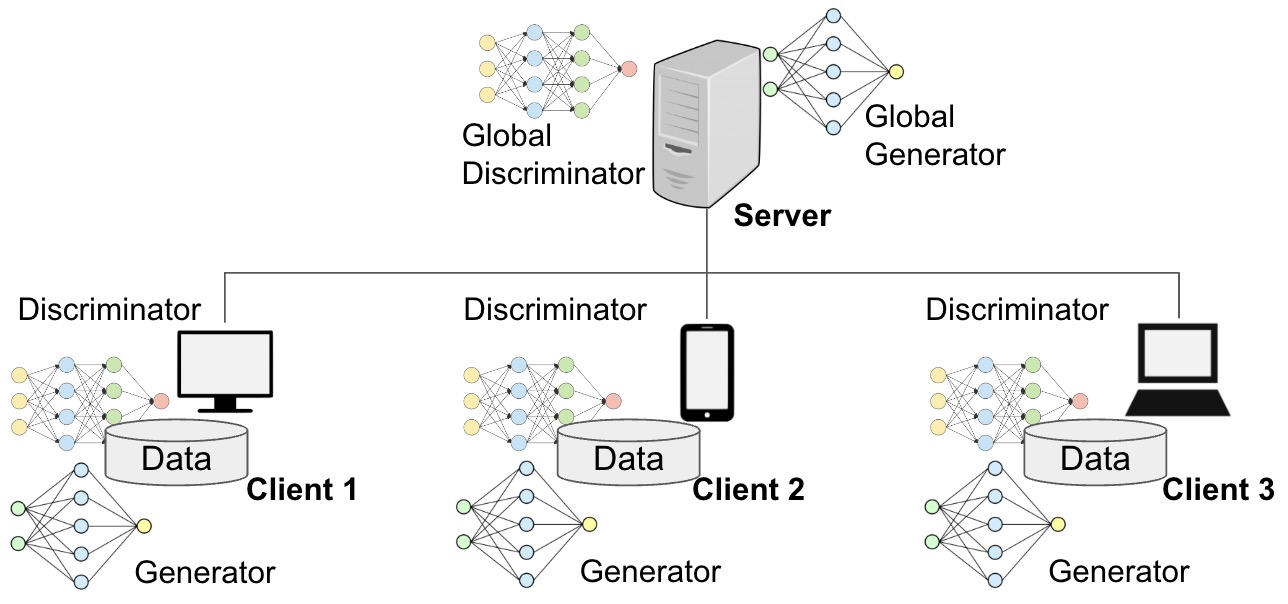}
}
\caption{Architecture of Distributed GAN. The left figure shows the multi-discriminator structure, where the server maintains a global generator and each client possesses a discriminator. The right figure shows the federated learning structure, where the server initializes both a global generator and discriminator and each client trains its local generator and discriminator on its dataset.}
\label{FIG-GAN}
\end{figure*}

\vspace{2mm}
\noindent\textbf{Deep learning-based method.}
% Zhao et al. \cite{zhao2023gtv} employed Generative Adversarial Networks (GAN) to develop a data synthetic model in their approach. In this setup, both each client and the server host a discriminator and a generator. The server's role involves training the generator to create comprehensive data records encompassing all features from every client. These generated data records are then partitioned according to each client and redistributed locally for individual client training. The intermediate logits (probability scores) derived from the clients' discriminators are transmitted to the server. The server aggregates these logits and forms a concatenated input, feeding it into its own discriminator for evaluation. To address the challenge of data imbalance, they utilize a conditional Generative Adversarial Network (GAN) \cite{xu2019modeling} to manage the generation of data by leveraging a conditional vector that controls the class of the generated information. However, this approach comes with inherent risks without appropriate perturbation. The server can potentially reconstruct client data by using the constructed conditional vector and the indices of selected training data. Even though a strategy is proposed to shuffle clients' datasets using the same random seed, there remains a significant vulnerability: if a single client collaborates with the server, the information can easily be exposed. 
The Distributed GAN framework has been employed for collaborative model training within a vertical setting. Two architectural variations, depicted in Figure \ref{FIG-GAN}, have been investigated. 
In the multi-discriminator structure, the server maintains a global generator, while each client possesses a discriminator; together, they collaboratively train a global generator. In the federated learning structure, the server initializes both a global generator and discriminator. Each client trains its local generator and discriminator on its dataset, updating the global models by aggregating parameter transformations from all clients. 

Jiang et al. \cite{jiang2023distributed} adopted a distributed GAN architecture consisting of one global generator on the server side and multiple discriminators on the client side. The global generator processes random latent features to generate synthetic data tailored for each local client. Meanwhile, the local discriminators are trained within each client's data to differentiate between real and synthetic data. DP perturbation is applied to the discriminator to protect data privacy. The proposed method successfully addressed the issue of attribute inconsistency and established connections among attribute columns across different clients. However, it does not account for the commonly occurring problem of data imbalance in tabular data. Consequently, its performance cannot be guaranteed when dealing with such imbalanced data distributions. Zhao et al. \cite{zhao2023gtv} solved the data imbalance issue by applying a conditional GAN. It manages the generation of data by leveraging a conditional vector that controls the class of the generated information. However, this approach comes with inherent risks without appropriate perturbation. The server can potentially reconstruct client data by using the constructed conditional vector and the indices of selected training data. Even though a strategy is proposed to shuffle clients' datasets using the same random seed, there remains a significant vulnerability: if a single client collaborates with the server, the information can easily be exposed. The proposed method doesn't employ differential privacy due to concerns about data utility. However, they did note that ensuring privacy could be achieved directly by introducing noise into the communicated intermediate result. 

\vspace{2mm}
\noindent\textit{Lessons learned and discussion.} The predominant emphasis of current research lies in the implementation of such data synthesis within a vertical framework. However, there's a notable scarcity in addressing specific challenges on tabular datasets, like handling imbalanced data distributions and managing the high dimensionality of attributes under differential privacy constraints. 
Furthermore, existing works commonly assume that all clients possess an identical set of records/individuals or can establish this through a secure set intersection protocol. This assumption presents a challenge since obtaining a substantial amount of identical user records from multiple parties for learning purposes might be impractical.

% Conversely, there exists an unexplored avenue that involves optimizing methods to fully leverage all available data records. This exploration holds promise in overcoming limitations imposed by insufficient shared user records while maximizing the potential of the available dataset.

\subsection{Horizontal data synthesis}
In a horizontal data distribution scenario, data records are spread across multiple nodes or servers, typically in a manner that divides data rows or entities across these nodes. For instance, various banks might hold the same categories of data about their customers, such as names, addresses, transaction histories, account balances, and so forth. Each bank can store this kind of information within its database, specifically for its customers, while maintaining separate entities from other banks. Specifically, within the context of horizontal data synthesis, each client holds a local dataset containing the same features related to different individuals. The particular challenge in synthesizing horizontal data is accurately estimating the overall data distribution, considering that each client's distribution may be skewed compared to the entire dataset.

\vspace{2mm}
\noindent\textbf{Statistical method.}
Su et al. \cite{su2016differentially} implemented multi-party data synthesis by collaboratively constructing a Bayesian Network. The parties and the server worked together to quantify correlations between attribute pairs and initialized a Bayesian network. Each party then sequentially updated it based on their local dataset. To manage noise and reduce communication costs, they constructed the search frontier of the network using only strongly correlated pairs. The sequential update method maximized the use of dependency information from previous parties, further reducing the number of candidate attribute pairs. Besides the inherent limitations of using Bayesian networks to generate synthetic data, it necessitates multiple uploads of information by each party, splitting the privacy budget into four pieces, thereby significantly impacting statistical accuracy. 

\vspace{2mm}
\noindent\textbf{Deep learning-based method.} Zhao et al. \cite{zhao2021fed} adopted the Federated learning collaborative training framework using CT-GAN \cite{xu2019modeling}. To efficiently encode the features, they suggested gathering private frequency information related to categorical attributes and parameters of a  Variational Gaussian Mixtures (VGM) model \cite{attias2013inferring} from individual users for each column. Following model initialization, this statistical information is reintegrated to compute weights for each client, aiding smoother convergence in scenarios of imbalanced data across different clients. Yet, the weight calculation relies on individual columns, disregarding the joint distribution of attributes. Also, it didn't explain how the concept of differential privacy could be used in the suggested method, even though it mentioned its potential application. It claimed that privacy is ensured by the transfer of statistical information rather than raw, original data. Fang et al. \cite{fang2022dp} incorporated differential privacy to CT-GAN by adding noise to the discriminator and deploying it in a federated setting as well. However, the feature encoding process is protected, which can still disclose some information about the original data. 
Besides the VGM model, the Variational Bayesian Gaussian Mixture Model (VB-GMM) \cite{corduneanu2001variational} is also considered \cite{duan2022ht} for modelling the distribution of the data column. In contrast to VGM, VB-GMM takes a Bayesian approach to estimate the model parameters and doesn't require specifying the number of clusters in advance. Duan et al. \cite{duan2022ht} implemented the data encoding process using Homomorphic Encryption \cite{aono2017privacy}. To prevent the discriminator from memorizing the private data during training, they introduced noise during the aggregation step of the discriminator. To well balance the privacy and data utility, Wang et al. \cite{wang2023atlas} proposed to adaptively adjust the noise scale to reduce the impact of gradient perturbation. Additionally, they utilize the discriminator to filter out more realistic synthetic data records during the data generation process.

\vspace{2mm}
\noindent\textit{Lessons learned and discussion.} %There are not many studies discussing using statistical models for distributed data synthesis. It is worth exploring more advanced methods to tackle the high-dimensional problem when collaboratively learning these statistical models. 
Few studies explore the use of statistical models for distributed data synthesis. Further research into advanced methods is needed to address high-dimensional challenges in collaborative learning of these models. For deep learning approaches, CT-GAN is used as the main tool in synthesizing tabular data through horizontal federated learning. To safeguard information during the feature encoding process, other privacy techniques like HE are employed. While it helps with privacy, it does increase communication and computation costs. Developing better ways to allocate privacy budgets and aggregate information could strike a better balance between privacy and data utility.

\section{Synthetic data evaluation}\label{Sec:evaluation}
\begin{figure}
    \centering
    \includegraphics[scale=0.2]{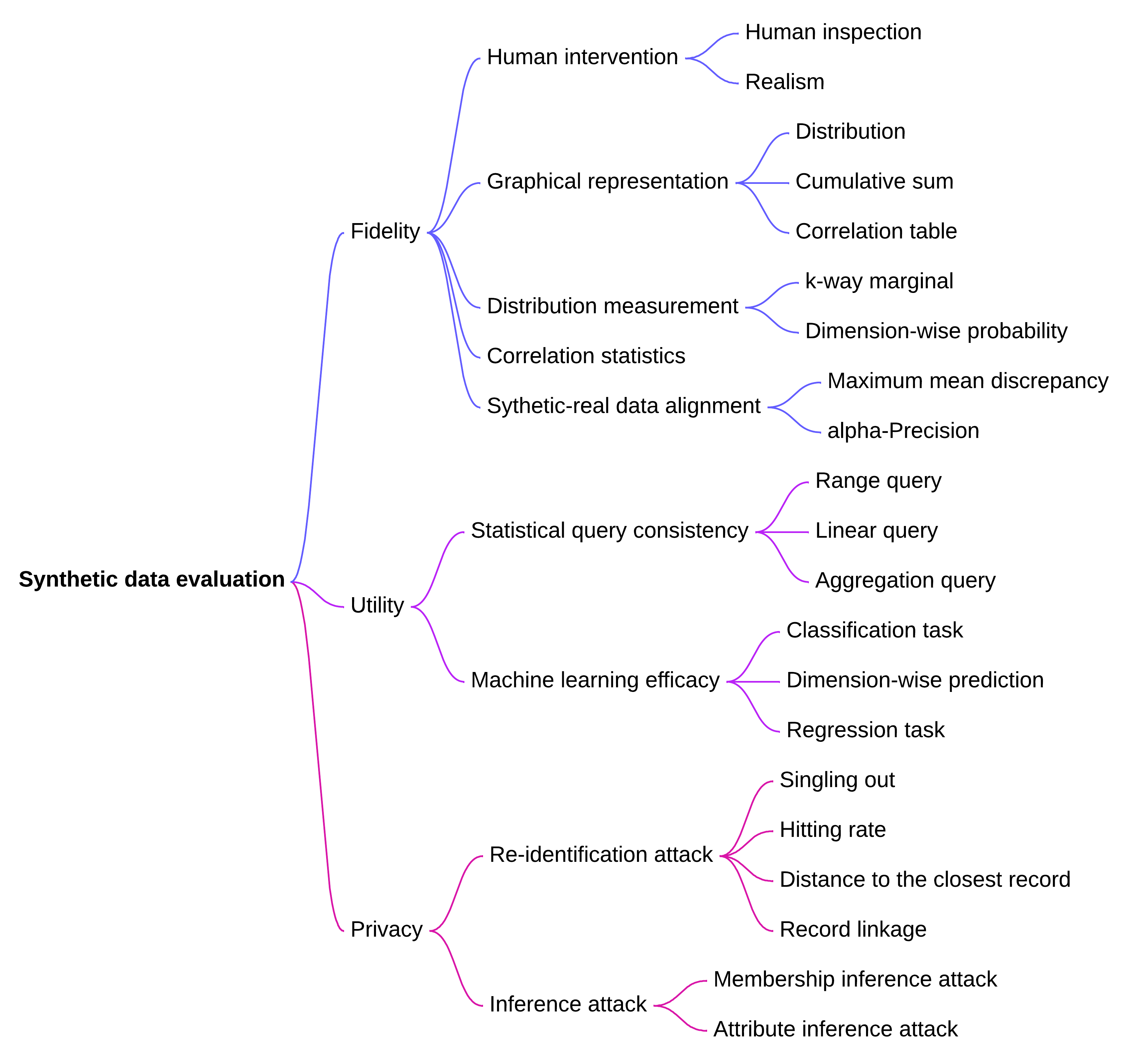}
    \caption{The taxonomy tree of the synthetic data evaluation methods}
    \label{fig:tt}
\end{figure}

Evaluating synthetic data is essential to ensure that it serves its intended purpose effectively while safeguarding sensitive information. 
Many evaluation techniques are designed for specific types of data or particular domains \cite{shmelkov2018good}. %In this section, we concentrate on evaluation methods tailored for tabular data shown in Fig. \ref{fig:tt}. %The taxonomy tree of the methods is shown in Fig. \ref{fig:tt}. 
In this section, we focus on evaluation methods for tabular data, with the taxonomy shown in Fig. \ref{fig:tt}.

% \begin{figure}
%     \centering
%     \includegraphics[scale=0.25]{Figures/eva.jpg}
%     \caption{The taxonomy tree of the synthetic data evaluation methods}
%     \label{fig:tt}
% \end{figure}

\subsection{Fidelity evaluation} 

Fidelity evaluation for synthetic data refers to the process of assessing how well the generated synthetic data matches the properties and patterns of the real data it aims to replicate. The primary goal is to ensure that the synthetic data is a reliable and accurate representation of the real data in terms of statistical, structural, and distributional characteristics.

\subsubsection{Human intervention.} 
This method is the most straightforward way to evaluate the fidelity of synthetic data. The metrics are concerned with how accurately the synthetic data replicates or aligns with the real-world.% or domain-specific data. 

\vspace{2mm}
\noindent\textbf{Human inspection}. This method requires the experts to judge whether the data is real or synthetic. Since the experts have professional knowledge, it helps them to identify fake data. For example, considering a clinical dataset, the fake data record might include a medical history with contradictory information, such as a patient having two mutually exclusive medical conditions or treatments that are incompatible \cite{beaulieu2019privacy}. 

\vspace{2mm}
\noindent\textbf{Realism}. Realism is similar to Human inspections, involving the verification of its alignment with domain-specific knowledge. It works by identifying a specific test, for each test, specifying the criteria that must be satisfied to classify the dataset as realistic. For instance, when evaluating a synthetic network traffic dataset \cite{fan2021dpnet}, if a flow describes normal user behavior and involves source or destination ports 80 (HTTP) or 443 (HTTPS), the transport protocol must be TCP.

\subsubsection{Graphical representation.} 
One method to evaluate the fidelity of synthetic data is through graphical representations. These visual assessments allow for easy verification of results and recognition of similar patterns between the real and synthetic data.

\vspace{2mm}
\noindent\textbf{Distribution.} It visualizes the distribution of each column or multiple columns of the real and synthetic data \cite{marin2022evaluating}, allowing for a visual assessment of how well the synthetic data matches the original data. A Q-Q (Quantile-Quantile) plot is an effective graphical method that compares the quantiles of the distribution of a synthetic dataset to the quantiles of the distribution of the original dataset. By plotting these quantiles against each other, the Q-Q plot helps identify whether the synthetic data preserves the statistical properties of the original data, highlighting any discrepancies between the two datasets.

\vspace{2mm}
\noindent\textbf{Cumulative sum.} Visualizing the cumulative sum of each column for real and synthetic data can reveal the similarity between their distributions as well. This method effectively demonstrates the distribution of both continuous and categorical values, facilitating a thorough comparison between the real and synthetic datasets. 

\vspace{2mm}
\noindent\textbf{Correlation table.} A correlation table displays the correlation coefficients between multiple variables in a dataset, with each cell representing the correlation between two variables. Comparing the correlation metrics can indicate how well the synthetic data captures the relationships between the columns \cite{goncalves2020generation}, thereby assessing the preservation of statistical dependencies in the synthetic data. These correlation coefficients can also be visually represented using a heatmap or a correlation matrix plot.

\subsubsection{Distribution measurement.} Distribution evaluation for synthetic data involves assessing how well the statistical properties of the synthetic data align with those of the real data. 

\vspace{2mm}
\noindent\textbf{$k$-way marginal.} $k$-way marginal, which refers to a marginal that consists of $k$ attributes, is the most commonly used metric to evaluate the statistical consistency between the synthetic dataset and the real datasets. As $k$ increases, %the metric assesses the correlations among attributes at higher orders. 
higher-order attribute correlations are measured. 
Various metrics can be used for quantifying the statistical errors, including $L_1$-distance, $L_2$-distance and $L_\infty$-distance. Besides, Hardt et al. \cite{hardt2012simple} utilized Kullback-Leibler (KL) divergence to measure entropy differences. While Gambs et al. \cite{gambs2021growing} employed Kolmogorov-Smirnov distance
for CDF comparisons. 
%Various distance metrics, including $L_1$-distance, $L_2$-distance and $L_\infty$-distance are used to quantify the statistical errors. Besides, Hardt et al. \cite{hardt2012simple} utilized Kullback-Leibler (KL) divergence, which measures the difference between the entropy of one probability distribution and the cross-entropy between two distributions. Gambs et al. \cite{gambs2021growing} employed Kolmogorov-Smirnov distance, defined as the maximum absolute difference between the two CDFs over all possible values of the variable, to estimate the fidelity of the distributions of the synthetic data compared to the original data. 
Du and Li \cite{du2024towards} proposed using Wasserstein distance to measure the distribution discrepancies. Additionally, total variation distance \cite{kotelnikov2023tabddpm} and the Kolmogorov-Smirnov test \cite{zhang2023mixed} are commonly used to assess distribution similarity for categorical and numerical attributes, respectively.

\vspace{2mm}
\noindent\textbf{Dimension-wise probability \cite{choi2017generating}}. Dimension-wise probability analysis is a fundamental validation technique used to assess a model's understanding of the distribution of data in each dimension. It is used for evaluating the model performance for binary variables. The process involves training the model on a real dataset and generating an equal number of synthetic samples. For every dimension $k$, it calculates the Bernoulli success probability ($p_k$) to estimate the likelihood of a specific outcome in that dimension. The $p_k$ values for each dimension in the real dataset are then compared to those in the synthetic samples. This comparison serves as a sanity check to determine whether the model has successfully captured the dimension-wise data distribution, ensuring the model's reliability.

\subsubsection{Correlation statistics.} Correlation evaluation for synthetic data focuses on assessing how well the relationships between variables in the synthetic data match those in the real data. Correlations are crucial for maintaining the structural integrity of the data, as they capture dependencies and associations between features. Researchers evaluate the similarity of pairwise relationships between variables in synthetic and real datasets by comparing correlation scores. Commonly used correlation measures include Theil’s uncertainty coefficient \cite{zhao2021ctab}, Pearson correlation \cite{zhang2023mixed}, and the correlation ratio \cite{kotelnikov2023tabddpm}. Additionally, Gambs et al. \cite{gambs2021growing} introduced the Mean Correlation Delta, a metric designed to quantify the differences in correlation coefficients between the correlation matrices of the real and synthetic datasets.

\subsubsection{Synthetic-real data alignment.} The group of metrics %assesses how well synthetic data aligns with the real data distribution.
evaluate how closely the synthetic data captures the patterns and characteristics of the real data, ensuring a strong alignment between the two. 

\vspace{2mm}
\noindent\textbf{Maximun mean discrepancy (MMD)}. MMD is a statistical measure that compares the similarity or difference between two sets of data points. Given two sets of data points, often denoted as $P$ and $Q$, first represent them in a higher-dimensional space using a kernel function. The choice of the kernel function affects how data points are mapped. MMD calculates the difference in mean values between the two sets in this higher-dimensional space. If the mean values are close, it suggests that the two sets are similar, and if they are significantly different, it suggests dissimilarity. 
In a recent study \cite{xu2018empirical}, MMD demonstrated most of the desired features of evaluation metrics, especially for GAN. 

\vspace{2mm}
\noindent\textbf{$\alpha$-Precision.} $\alpha$-Precision is proposed in \cite{alaa2022faithful}. It is an enhanced version of the conventional Precision metric. While Precision checks how often generated data matches real data, $\alpha$-Precision goes further by focusing on a specific portion of the real data distribution. It evaluates whether a synthetic data point falls within the most important or central part of the real data distribution, defined by a parameter $\alpha$. In doing so, it measures the fidelity of synthetic data, ensuring that it captures the core characteristics of the real data, rather than just any random part. The core part refers to the region of the data distribution where the majority of real data points are concentrated, representing the most significant or typical patterns within the data.

\subsubsection{Discussion.}
%Human evaluation methods could be employed for practical purposes, though there may be some drawbacks, such as being expensive or time-consuming. It is rarely utilized in the academic literature to assess the quality of synthetic data. 
Human evaluation methods, though practical, are rarely used in academic literature to assess synthetic data quality due to their cost and time demands.
%Graphical evaluation provides a visual comparison of datasets, making it easier to assess the similarity between real and synthetic data. However, when dealing with datasets containing hundreds or more variables, this approach can become resource-intensive and challenging to present comprehensively. 
Graphical evaluation visually compares real and synthetic data, aiding similarity assessment, but becomes resource-intensive and complex with large, high-dimensional datasets. The simplest method to evaluate the fidelity of synthetic data is by using basic statistics such as mean, median, and variance \cite{charest2011can}. The closer these statistics are between the real and synthetic data, the more similar the datasets are considered. However, this approach may not capture complex relationships within the data. Francis Anscombe demonstrated this with his famous ``Anscombe's quartet" \cite{anscombe1973graphs}, where four datasets had nearly identical basic descriptive statistics but vastly different distributions. This illustrates the need for more sophisticated and in-depth statistical analysis to evaluate synthetic data.% accurately. 

\subsection{Utility evaluation} Utility evaluation is the process of assessing how useful synthetic data is for specific tasks or applications, typically by comparing its performance with that of real data. The goal is to ensure that synthetic data preserves enough meaningful information from the original data to be effectively used in analysis, decision-making, or machine learning models.

\subsubsection{Statistical query consistency.} One common metric used to evaluate the utility of synthetic data is through statistical queries. The focus is on assessing how well the synthetic data preserves the statistical properties of the real data when responding to specific queries. This process determines whether the synthetic data retains enough accuracy and relevance to be used for meaningful analysis, ensuring its utility for downstream tasks.

\vspace{2mm}
\noindent\textbf{Range query.} A range query is a type of database query that retrieves all records where the values in a specific field fall within a specified range. This is particularly useful for querying numerical, date, or other ordered data types. The range query can involve one or more attributes. For instance, it might inquire about the percentage of data records with ages between $40$ and $60$ and salaries between $1000$ and $2000$ dollars. Range queries are commonly used in evaluating synthetic data \cite{li2014differentially,liew2022pearl}. 

\vspace{2mm}
\noindent\textbf{Linear query.} Linear query, in the context of synthetic data generation or data analysis, typically refers to a query that can be expressed as a linear combination of features or attributes in a dataset \cite{xu2017dppro}. Notably, in the work of Acs et al. \cite{acs2012differentially}, a linear query is defined as a predicate function that calculates the count of instances in the dataset that meet the specified predicate criteria.

\vspace{2mm}
\noindent\textbf{Aggregation query.} An aggregation query is a query that processes and summarizes data to produce a single result or a set of results based on grouped data. These queries are commonly used to compute aggregate values like sums, averages, counts, minimums, maximums, and other statistical metrics. Fan et al. \cite{fan2020relational} ran the same aggregation queries with aggregate functions, such as count, average and sum, on both original and synthetic datasets, and then measured the relative error of the results.

\subsubsection{Machine learning efficacy.} Machine learning-based evaluation involves using machine learning models to assess the utility of synthetic data. This evaluation can be done by performing specific predictive tasks, %such as classification, regression, clustering, or any other predictive task, 
and comparing the performance of these models when trained on real data versus synthetic data. 

\vspace{2mm}
\noindent\textbf{Classification task.} Classification tasks are commonly employed to assess the preservation of attribute correlations in relation to attribute prediction. 
This involves training classifiers on both the original and synthetic datasets and then testing them on a shared test set. Performance is usually compared using metrics like misclassification rate or F1 score. 
%The utility of this assessment is determined by training two classifiers, one on the original dataset and the other on the synthetic dataset. Subsequently, both classifiers are tested using the same test dataset sampled from the original dataset. The comparison is based on the misclassification rate of each method or F1 score. 
Besides, Bindschaedler et al. \cite{bindschaedler2017plausible} introduced the agreement rate, 
which measures the percentage of identical predictions. 
%which is defined as the percentage of instances where two classifiers produce identical predictions, irrespective of the correctness of those predictions, to evaluate the similarity of two classifiers. 
Du and Li \cite{du2024towards} proposed Machine Learning Affinity to measure the relative performance discrepancy across models. Ideally, the classifier trained on the synthetic dataset should exhibit classification performance similar to that of the one trained on the original dataset. 

\vspace{2mm}
\noindent\textbf{Dimension-wise prediction.} Dimension-wise prediction was proposed by Choi et al. \cite{choi2017generating} to quantify the binary variables. This approach seamlessly extends to handling multiple variables. Differing slightly from traditional classification methods, this approach allows the flexibility of selecting any attribute as the label for prediction.

\vspace{2mm}
\noindent\textbf{Regression task.} The regression task has been employed to assess synthetic data as well in the literature \cite{gambs2021growing}. This involved training two predictors separately on synthetic and real datasets. Subsequently, these predictors were utilized to forecast values for a specific attribute within the test dataset. The ensuing step involved a comparative analysis of the predicted values generated by the two distinct predictors.

\subsubsection{Discussion.}
Query-based metrics are typically computationally efficient and do not require complex models or large-scale processing. However, they only evaluate specific statistics or properties being queried, which means they may not fully capture the performance of synthetic data in task-specific scenarios. Machine learning-based evaluation directly assesses how well synthetic data supports specific tasks, providing clear and quantitative metrics for comparison. However, training and evaluating machine learning models can be computationally intensive, particularly with large datasets or complex models such as deep neural networks. Besides, it is well known that tabular data demonstrates varying performance across different machine learning models, with no single model consistently achieving optimal results on all datasets \cite{du2024towards}. Consequently, making fair comparisons using machine learning-based methods becomes challenging and needs careful design.

\subsection{Privacy evaluation}
%A synthetic dataset serves as a means to share sensitive information in a private manner, preserving the global statistical properties of the original data without revealing sensitive details. Differential privacy techniques are employed to fortify such privacy assurances. Assessing the residual privacy risks becomes a critical concern. Although comprehensive studies quantifying the privacy risks associated with differential privacy synthetic data are limited, evaluations based on specific attacks offer valuable insights into the privacy levels offered by these private data synthesis methods. 
Privacy evaluation for synthetic data assesses how well it preserves the privacy of individuals in the original dataset. Even with differential privacy techniques, residual risks may persist, making accurate risk quantification essential. Limited research exists on privacy risks in differentially private synthetic data, but attack-based assessments offer valuable insights into protection levels.
The generic privacy risk evaluation assumes the synthetic data generation process is a black box. In this scenario, the attacker can only access the generated synthetic data and attempt to deduce sensitive information by executing various attacks on the synthetic dataset. The attacker might possess some background knowledge about the individuals involved. 

\subsubsection{Re-identification attack.}  One group of evaluation aims to quantify the risk associated with re-identifying the original data records.

\vspace{2mm}
\noindent\textbf{Singling out.} The intuition behind this method is that individual data records can be isolated if their attributes or attribute connections are unique in the synthetic data. In the case of numerical attributes, Xue et al. \cite{giomi2022unified} employed the minimum and maximum values of each attribute to establish predicates based on values falling below the minimum or exceeding the maximum, enabling the identification of outliers. 

\vspace{2mm}
\noindent\textbf{Hitting rate.} The hitting rate measures the number of records in the original dataset that a synthetic record can match. Two records are considered similar only if all categorical attribute values are identical, and the discrepancy between numerical attribute values falls within a defined threshold \cite{fan2020relational}. 

\vspace{2mm}
\noindent\textbf{Distance to the closest record.} This metric assesses the proximity of a synthetic data record to the nearest real data record. A distance of $0$ indicates that the synthetic data perfectly replicates a real data record, potentially revealing actual information. Conversely, a greater distance signifies stronger privacy protection, indicating a larger divergence from the real data \cite{lu2019empirical}.

\vspace{2mm}
\noindent\textbf{Record linkage.} The assessment of privacy risk relies on the success rate of identifying connections between synthetic data records and their corresponding original data records. Various record linkage methods, including probabilistic \cite{jaro1989advances} and distance-based approaches \cite{mateo2004outlier}, can be employed for this evaluation. Lu et al. \cite{lu2019empirical} utilized the Python Record Linkage Toolkit to compare record pairs.

\subsubsection{Inference attack.} Another group of evaluations seeks to measure the synthetic dataset's capacity to deduce information about the membership and specific sensitive attributes.

\vspace{2mm}
\noindent\textbf{Membership inference attack.} 
A membership inference attack \cite{shokri2017membership} on synthetic data aims to determine if a particular individual's data contributed to training the synthetic generative model. Stadler et al. \cite{stadler2022synthetic} attempted to attack synthetic data by utilizing handcrafted features from the synthetic data distribution to train shadow models. Hyeong et al. \cite{hyeong2022empirical} estimated the likelihood of target records using density estimation and applied this as a confidence score for membership inference. Gambs et al. \cite{gambs2021growing} introduced the Monte Carlo attack, which calculates the count of synthetic data records neighbors of the queried data record, thereby inferring its potential presence in the training dataset. 

\vspace{2mm}
\noindent\textbf{Attribute inference attack.} Attribute inference operates under the assumption that the attacker possesses knowledge about certain attributes of the victim. Leveraging this background information and the synthetic dataset, Giomi et al. \cite{giomi2022unified} deduced sensitive attributes by searching for the closest data record containing the known attributes. The sensitive attributes in the nearest data record constitute the attacker's estimated information. Stadler et al. \cite{stadler2020synthetic} performed the attribute attack by training a machine learning model on the synthetic dataset and subsequently predicting the sensitive attribute of the victim based on the known attribute values. 
Annamalai et al. \cite{annamalai2023linear} proposed a linear reconstruction attack method that generates a series of marginal queries to infer the sensitive attribute by minimizing the query error based on the known victim attributes.

\subsubsection{Discussion.}
Beyond generic attacks, if an attacker gains access to the data generation model, traditional membership \cite{hu2022membership} and attribute \cite{zhang2023survey} attacks for machine learning could be employed to assess privacy risks. While most of these attacks traditionally focus on image data, exploring effective methods to evaluate the risks of tabular data utilizing such attacks could be an intriguing study area. Besides, developing a unified framework to quantify the privacy risk could benefit society. It can inform the development of privacy regulations and policies by providing a structured approach to assessing and mitigating risks, aiding in establishing effective guidelines and standards.

\section{Research gaps and future directions}\label{frd}
% This section explores the research gaps and future research directions identified through the review of existing methods. 

This section explores current research gaps, highlights areas where current solutions fall short, and identifies potential future research directions. It focuses on design, privacy, evaluation methods, and emerging application scenarios that present promising opportunities for future research advancements

% \subsection{Research gaps}
% By analyzing the strengths and limitations of the approaches surveyed, we highlight areas where current solutions fall short, pointing to gaps in the literature.

\subsection{Tailored DL Methods for Tabular Data Synthesis.} Deep learning-based data synthesis methods offer a powerful and flexible approach to generating synthetic data. These methods can efficiently handle high-dimensional datasets and scale well. While they show great potential for tabular data synthesis, most existing studies have overlooked the unique characteristics of tabular data. These include mixed data types, unbalanced data distributions, and column interdependencies. Some strategies have been proposed to address these challenges \cite{tantipongpipat2021differentially,zhao2023gtv}. For instance, autoencoders are commonly used to handle mixed data types, while some deep learning models introduce conditional GANs to manage unbalanced data distributions. However, research in this area is still in its early stages, and more effective strategies need to be explored to enhance the capability of deep learning models for tabular data synthesis. These strategies may include the structural design of neural networks and the integration of specific statistical components to capture the distribution and dependencies of the dataset.

\subsection{Privacy risk evaluation.} %As discussed in Section \ref{Sec:evaluation}, the majority of research has to date focused on evaluating the utility of generated synthetic data. However, 
While DP offers a provable level of privacy protection, it remains unclear how much privacy risk remains at a fixed privacy level $\epsilon$. In other words, it is uncertain how large a privacy budget can provide sufficient privacy protection. A study \cite{stadler2020synthetic} suggested that generative models trained without privacy-preserving techniques (non-private generative models) offer limited protection against inference attacks.
Furthermore, the researchers discovered that training generative models with differential privacy did not significantly enhance protection against inference attacks. Another recent study \cite{hyeong2022empirical} found that synthetic datasets generated by DP-GAN exhibit better resistance to black-box attacks, but white-box attackers can still accurately infer membership. This raises questions about the effectiveness of other synthetic data generation models, such as marginal-based methods, in safeguarding privacy. We may be interested in assessing the effectiveness of privacy protection against various attacks or exploring methods to strike a balance between preserving privacy and maximizing utility. 

\subsection{Differentially private tabular data synthesis with the diffusion models.} %Diffusion models are a class of generative models known for their ability to produce high-quality, diverse images. They become prevalent in the area of computer vision, particularly in tasks involving image generation and manipulation. 
Diffusion models have recently been explored for tabular data synthesis since their flexible probabilistic framework can effectively model complex data distributions and capture the dependencies. 
Several works \citep{zhang2023mixed,kotelnikov2023tabddpm,suh2023autodiff,sattarov2023findiff,lee2023codi} addressed the challenges of synthesizing tabular data with mixed data types and varied distributions. Yang et al. \citep{yang2024balanced} focused on ensuring fairness in synthetic data generation using diffusion models. Villaizan et al. \citep{villaizan2024diffusion} and Jolicoeur et al. \citep{jolicoeur2024generating} explored generating and imputing tabular data with diffusion models. Additionally, some studies propose using federated learning for tabular data generation while maintaining data privacy. Sattarov et al. \citep{sattarov2024fedtabdiff} combined denoising diffusion probabilistic models with federated learning to enhance data privacy in the horizontal scenario. Meanwhile, Shankar et al. \citep{shankar2024silofuse} investigated data synthesis using diffusion models with vertically partitioned data. However, the use of diffusion models for synthetic data generation has not been explored in conjunction with differential privacy, either in centralized or distributed settings. Investigating the challenges and balancing privacy and utility when integrating differential privacy into the diffusion model, which offers robust protection during the data synthesis process, would be a valuable area of research.

% \subsection{Future research directions}
% By assessing the current state of the field, we identify several key areas that offer promising opportunities for future research development.

\subsection{Disparity effect of DP.} DP noise often leads to negative values and introduces inconsistencies within marginal distributions. Various post-processing techniques are commonly employed to enhance the accuracy of statistical results \cite{9264723}. %For instance, a common approach to address negative values is to round them to zero and normalize the marginal distributions. To ensure consistency across marginal distributions, the widely adopted method uses weighted averages to minimize overall variance. 
However, it is important to note that while post-processing techniques can improve statistical accuracy, they may introduce some bias in the data \cite{zhu2021bias,yang2024privacy}. The bias introduced during the post-processing of noisy data can propagate and become more pronounced in downstream tasks using synthetic data. This observation is demonstrated in the research conducted by Ganev et al. \cite{ganev2022robin}. Their recent study empirically illustrates the differential privacy's varying impact on synthetic datasets. Particularly, classifying tasks on the private synthetic dataset influences the gaps between the majority and minority subgroups. PrivBayes reduces this disparity, while another GAN-based model exacerbates it. Research on the disparity effects of applying differential privacy in synthetic data generation is still limited. %Exploring this topic further, investigating the factors that contribute to these effects, and developing mitigation to address them are crucial areas of study. 
Further exploration of contributing factors and developing mitigation strategies are crucial areas of study.

\subsection{Multi-party data synthesis with different users.} In distributed data synthesis scenarios, particularly when datasets are vertically partitioned, the literature often assumes that the two parties have precisely the same user groups or that overlapping users can be identified through Private Set Intersection (PSI) techniques \cite{morales2023private}. The synthetic data learning process is then conducted on these overlapping users. However, it is rare for different organizations to have the same users in practice. A significant issue arises when the size of the overlapping user group is small. On the one hand, it becomes challenging to train an effective generative model; on the other hand, it leads to substantial data resource wastage, as most data cannot be utilized for model learning. Therefore, developing efficient collaborative data generation methods that consider scenarios where parties hold different user groups is promising and essential. Such methods should utilize all available data to improve the quality of the generated models. 

%CuTS: Customizable Tabular Synthetic Data Generation

\subsection{Differentially private tabular data synthesis with LLM.} The popularity of Large Language Models (LLMs) has surged in recent years, driven by their remarkable capabilities in natural language processing. While initially designed for text-based tasks, LLMs have been adapted to work with structured data like tables \citep{fang2024large,bordt2024elephants}. Borisov et al. \citep{borisov2022language} proposed using pre-trained LLMs for synthesizing tabular data in a non-private setting.%, leveraging their ability to learn and replicate complex patterns, handle structured and high-dimensional data, and preserve statistical properties. 
By treating each row as a sequence of tokens, LLMs can generate tabular data like generating text, showing promising results that outperform GANs in synthesizing tabular data. However, when differential privacy is applied using DPSGD, Tran and Xiong \citep{tran2024differentially} demonstrated that traditional DP fine-tuned LLMs struggle to generate tabular data with format compliance due to the injected noise. To address this, they proposed a two-phase fine-tuning method that first learns the format and then fine-tunes the model to capture the feature distributions and dependencies of the dataset. %The use of LLMs for generating synthetic data, especially under differential privacy constraints, is still in its infancy, but it represents a promising and exciting area for further exploration. 
The use of LLMs for generating synthetic data, especially under differential privacy constraints, is still in its early stages. Since LLMs are primarily designed for unstructured text, adapting them to generate structured or tabular data presents significant challenges, especially when it comes to capturing the complex relationships between columns and rows. Additionally, achieving the right balance between ensuring privacy and maintaining the utility of the generated data remains an ongoing challenge. % However, this remains a promising and exciting area for further exploration.

\subsection{Multi-relational data synthesis} 
Research on tabular data synthesis primarily focuses on single-table scenarios, while in practice, data is often stored in multiple interconnected tables as relational databases. The main challenge in multi-table synthesis is capturing the long-range dependencies caused by foreign-key relationships. Few studies have explored this area. For instance, Mami et al. \citep{mami2022generating} employed graph variational autoencoders to model the synthesis process, while Patki et al. \citep{patki2016synthetic} used Gaussian copulas to capture parent-child relationships. Another work \citep{cai2023privlava} leveraged the controlled generation capabilities of diffusion models, utilizing clustering labels as intermediaries to model relationships between tables. To the best of our knowledge, only one study addresses the synthesis of relational data with foreign keys under differential privacy. The key idea is to model the data distribution using graphical models, incorporating latent variables to capture inter-relational correlations induced by foreign keys. Calibrated noise is injected into the model training algorithm to ensure differential privacy.
Incorporating differential privacy into multi-relational data synthesis is still an open problem, particularly for large-scale databases with complex dependencies among hundreds of interconnected tables. DP noise makes it difficult to capture cross-table correlations, and distributing the privacy budget effectively across interconnected tables while still ensuring the privacy of sensitive information presents another significant challenge.

% \vspace{2mm}
% \noindent\textbf{Targeted data synthesis} 

\section{Conclusion} \label{con}
In this paper, we conducted a comprehensive review of existing DP methods for synthesizing tabular data, a widely used data type in finance, healthcare, and other domains. We categorized these methods into statistical methods and deep learning-based approaches based on the data generation models, discussed in both centralized and decentralized settings. We examined and compared the methods in each category, highlighting their strengths and weaknesses regarding utility, privacy, and computational complexity. Furthermore, we provided a comprehensive overview of evaluation techniques for data synthesis, including fidelity, utility and privacy. From this analysis and discussion, we identified the research gaps and several potential directions for future research.


% \begin{acks}
% \end{acks}

%%
%% The next two lines define the bibliography style to be used and
\bibliographystyle{ACM-Reference-Format}
\bibliography{sample-base}


\end{document}